\pdfoutput=1

\documentclass{article}

\usepackage[T1]{fontenc}
\usepackage{microtype}
\usepackage{graphicx}
\usepackage{subcaption}
\usepackage{booktabs}
\usepackage{mathtools}

\usepackage[pagebackref=true]{hyperref}
\renewcommand*\backref[1]{\ifx#1\relax \else (Cited on #1) \fi}

\usepackage[accepted]{icml2026}

\usepackage{url}
\usepackage{xurl}
\usepackage{array}
\usepackage{xcolor}
\usepackage{colortbl}
\usepackage{caption}
\usepackage{enumerate}
\usepackage{listings}
\usepackage{mdframed}
\usepackage[most]{tcolorbox}
\tcbuselibrary{theorems,skins,breakable,listings}

\lstdefinestyle{prompt}{%
  basicstyle=\fontfamily{cmtt}\selectfont\footnotesize,
  breaklines=true,
  breakatwhitespace=false,
  columns=fullflexible,
  keepspaces=true,
  showstringspaces=false,
  postbreak=\mbox{\textcolor{gray}{$\hookrightarrow$}\space}
}

\lstdefinestyle{promptstyle}{style=prompt}
\usepackage{soul}
\usepackage{enumitem}
\usepackage{changepage}
\setlist{nosep,leftmargin=*}
\usepackage{tikz}
\usepackage{pgfplots}
\pgfplotsset{compat=1.18}
\usepackage{lineno}

\definecolor{darkblue}{rgb}{0, 0, 0.5}
\hypersetup{colorlinks=true, citecolor=darkblue, linkcolor=darkblue, urlcolor=darkblue}
\usepackage[capitalize,noabbrev,nameinlink]{cleveref}

\newcommand{\pmark}[1]{\textsuperscript{\textcolor{teal}{\ensuremath{#1}}}}
\newcommand{\nmark}[1]{\textsuperscript{\textcolor{red!70!black}{\ensuremath{#1}}}}
\newcommand{\softtoprule}{\arrayrulecolor{black!55}\specialrule{0.45pt}{0pt}{2pt}\arrayrulecolor{black}}
\definecolor{takeawaygreen}{RGB}{116,154,114}
\definecolor{takeawaylight}{RGB}{242,248,238}
\newtcolorbox{takeawaybox}{
  enhanced jigsaw,
  breakable,
  frame hidden,
  colback=takeawaylight,
  boxrule=0pt,
  arc=1pt,
  left=5pt,
  right=5pt,
  top=4pt,
  bottom=4pt,
  before skip=0.85\baselineskip,
  after skip=0.65\baselineskip,
  boxsep=2pt,
  overlay={
    \draw[takeawaygreen, line width=1.6pt]
      ([yshift=-1pt]frame.north west) -- ([yshift=1pt]frame.south west);
  },
  fontupper=\small
}

\icmltitlerunning{Is this Citation on Point?}

\begin{document}

\twocolumn[
  \icmltitle{Is this Citation on Point?}

  \begin{icmlauthorlist}
    \icmlauthor{Apurv Verma}{bb,njit}
  \end{icmlauthorlist}

  \icmlaffiliation{bb}{Bloomberg, New York, NY, USA}
  \icmlaffiliation{njit}{New Jersey Institute of Technology, Newark, NJ, USA}

  \icmlcorrespondingauthor{Apurv Verma}{averma239@bloomberg.net, av787@njit.edu}

  \icmlkeywords{Legal NLP, Citation Verification, LLM Evaluation, Hallucination Detection, Legal AI Safety}

  \vskip 0.3in
]

\printAffiliationsAndNotice{}

\begin{abstract}
In 2023, a New York judge sanctioned two attorneys in \emph{Mata v. Avianca} for filing a brief containing hallucinated citations generated by ChatGPT. These failures can be largely caught by database lookups. The harder problem is detecting citations that point to real cases but do not support the propositions for which they are offered, a failure mode that existing evaluations of LLMs for legal use cases largely overlook. In this paper, we study proposition-level citation support verification through controlled perturbations of real legal citations obtained from two legal corpora. We either replace the cited case or change only the pinpoint page within the same case. Our evaluation of fourteen model configurations on the resulting examples shows that models catch $93$--$100\%$ of wrong-case corruptions, but they only catch $37$--$61\%$ of wrong-pinpoint corruptions on court opinions and $52$--$83\%$ on legal briefs. When models fail to catch wrong-pinpoint corruptions, they accept the citation based on topical overlap rather than page-level support. Larger-scale models and extended reasoning narrow the gap but do not close it: GPT-5.4 with high reasoning effort still misses $40\%$ of pinpoint mismatches on court opinions and $18\%$ on briefs. Prompting the model to verify support at the cited page improves recall, but it also raises the false positive rate. These results show that citation misrepresentations remain difficult to detect automatically. Recognizing the right legal topic and verifying support for the cited proposition are distinct capabilities, and current models conflate them.
\par\vspace{0.5em}
{\noindent\textbf{Keywords:} \textit{Legal NLP; legal citation verification; LLM evaluation; hallucination detection; guardrails; AI safety.}\par}
\end{abstract}

\section{Introduction}
\label{sec:introduction}
\begin{figure}[t]
\centering
\includegraphics[width=\columnwidth]{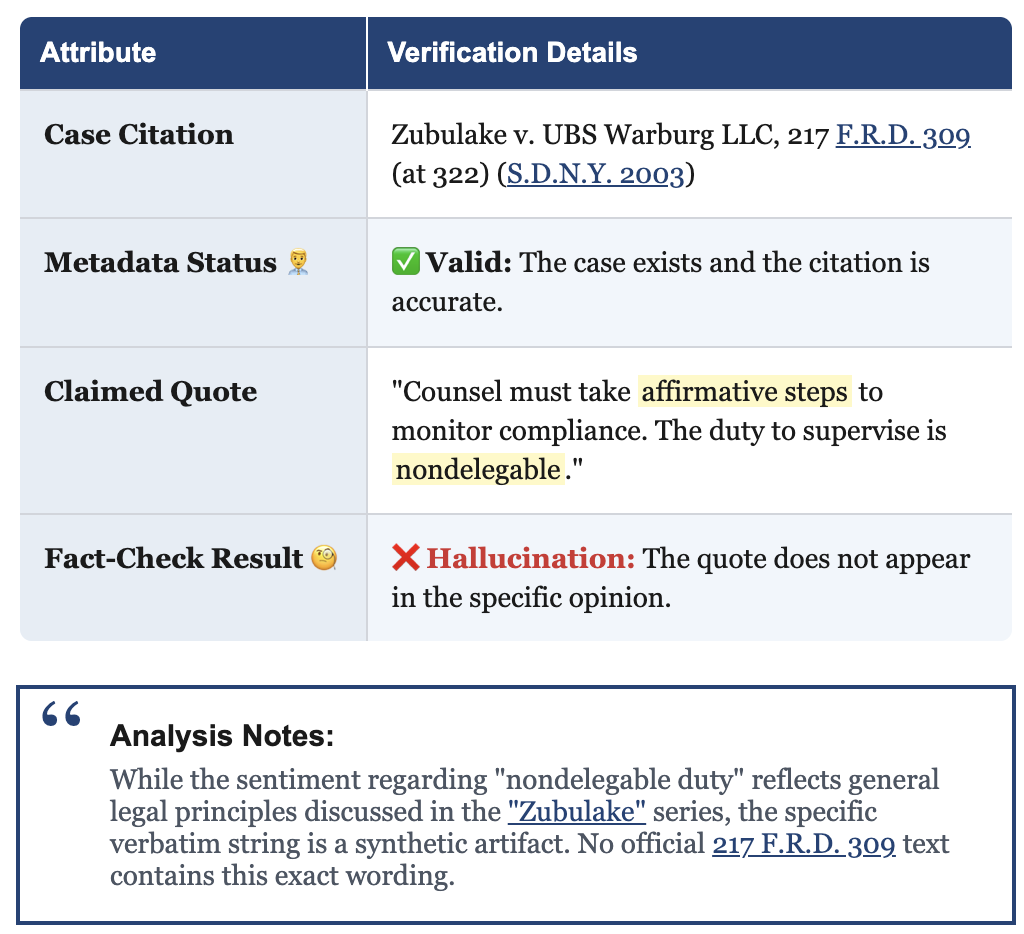}
\caption{A real citation error from a court filing \citep{AIHalluc47:online}. The case \textit{Zubulake v. UBS Warburg} exists and is correctly cited, but the quoted holding does not appear in the opinion. This type of error, where the citation is real but the attributed proposition is not supported by the opinion, is harder to detect than a non-existent citation.}
\label{fig:real-example}
\end{figure}

Legal citations do more than name an authority: they represent that the cited authority supports the claim being made, and attorneys are professionally obligated to verify this representation before filing under Model Rules 3.3 and 1.1 \citep{modelrules2020}. \Cref{fig:real-example} shows a real court filing where the case and reporter are correct, but the cited page does not support the quoted proposition. In practice, \textit{citechecking} involves both checking citation form and confirming source support \citep{keele2012librarians}. This paper studies that second task.

\begin{figure*}[t]
    \centering
    \includegraphics[width=\textwidth]{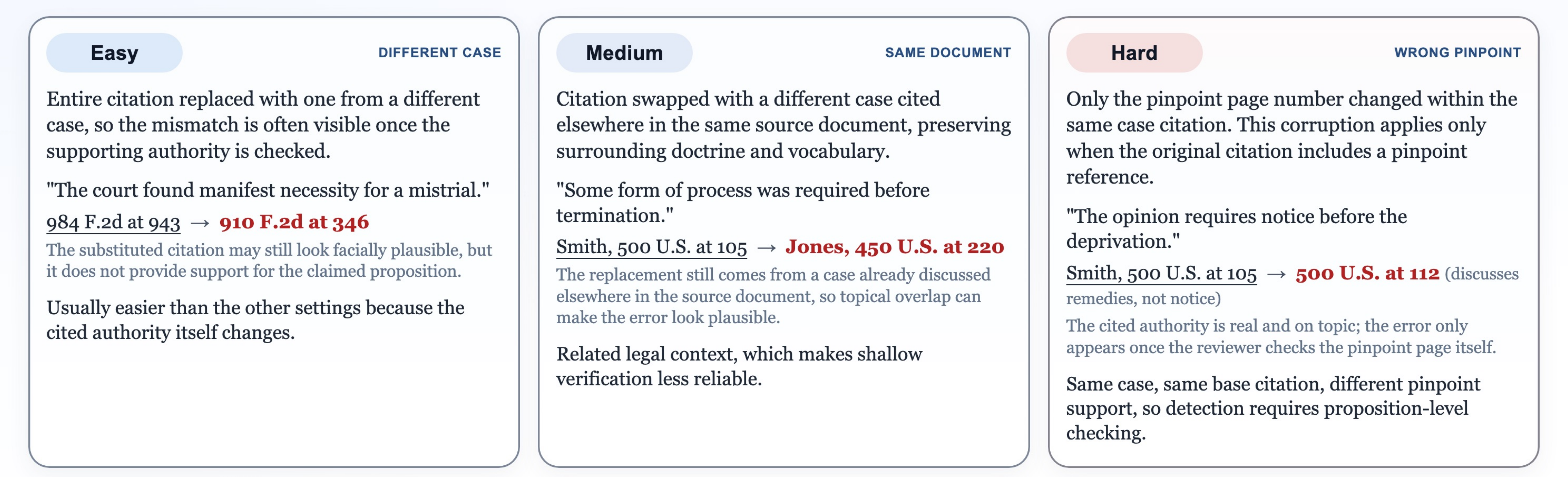}
    \caption{Progression of citation corruptions from easy to hard. Easy corruptions replace a citation with a different case. Medium corruptions substitute a citation with another case cited elsewhere in the same source document. Hard corruptions alter only the pinpoint page within the same case citation, so this setting applies only when the original citation includes a pinpoint reference. This mirrors real citechecking. Spotting a different authority is often easier than verifying that a cited pinpoint page supports the proposition.}
    \label{fig:corruption-examples}
\end{figure*}

That task becomes harder when LLMs are used for drafting since they produce plausible briefs with dozens of citations in seconds, increasing the volume of support that must be checked. Fluency and superficial plausibility encourage overreliance on LLM-generated legal text even when its claims lack support \citep{mik2023caveat}. We ask a narrow question: \textbf{Can current LLMs detect when a legal citation does not support the proposition for which it is offered?}

Following \citet{Dahl_2024}, citation failures come in two forms: (1)~\textit{fabrication}, where the cited authority does not exist, and (2)~\textit{misrepresentation}, where the authority exists but does not support the proposition for which it is cited. Fabricated cases have appeared in court filings \citep{matacase}, and fabricated citation URLs persist across frontier models even in general research settings \citep{rao2026detecting,xu2026ghostcite}, though database lookups, retrieval grounding and agentic self-correction have reduced their frequency \citep{magesh2025hallucination}. Misrepresentations are much harder to detect. Here, the case, the reporter, and the pinpoint page are all real, but the cited authority does not support the proposition. Unlike open-domain factuality evaluation, where topical relevance to a source usually implies support or contradiction, legal propositions can share doctrinal vocabulary and factual context with a cited authority that does not actually support them. One reason for this may be because legal arguments often proceed by analogy rather than direct entailment, and two passages can discuss the same doctrine while supporting different propositions \citep{huang2025survey,min-etal-2023-factscore,Sarol_Ming_Radhakrishna_Schneider_Kilicoglu_2024,michaels2016holding_dicta, 10.1093/acprof:oso/9780199756148.003.0009,10.1093/acprof:oso/9780199756148.003.0008,schauer2017analogy}. As fabrications become easier to catch, this second failure mode becomes more consequential, not less. That is our focus in this paper.

Citations in a legal text serve various functions. Some may record the procedural history of a case, while others may provide direct doctrinal support, or supply background. We focus on \emph{substantive} citations---those providing direct support---because errors in such citations change the meaning of the legal argument being made (\cref{sec:taxonomy} and \cref{appendix:legal_background} cover relevant legal background).
In U.S.\ legal writing, citations follow the Bluebook citation format \citep{bluebook2020}. At the coarsest level, we can swap the cited case entirely; when the citation includes a pinpoint page reference, we can also change only the page within the same case. This produces controlled errors at three difficulty levels (\cref{fig:corruption-examples}; details in \cref{sec:corruption-strategies}).
We do not claim that these corruptions mirror the exact citation errors that one might expect to see in real-world LLM drafting. However, they provide a useful controlled probe for checking if models can verify propositional support.

We organize the study around three research questions:
\begin{itemize}
    \item \textbf{RQ1.} How well do LLMs detect citation support errors across difficulty levels?
    \item \textbf{RQ2.} Do larger or newer models (e.g., GPT-5 vs.\ GPT-4o) perform better?
    \item \textbf{RQ3.} Does performance vary across document types (court opinions vs.\ briefs)?
\end{itemize}

Across fourteen frontier model configurations from three families and two legal datasets, we see that models nearly saturate on Easy corruptions ($93$--$100\%$ recall on both datasets) but struggle on Hard ones, where only the pinpoint page changes: recall falls to $37$--$61\%$ on court opinions and $52$--$83\%$ on briefs. GPT-5.4 with extended reasoning (chain-of-thought) gains $11$~pp (percentage points) on court opinions and $18$~pp on briefs, and still misses $40\%$ of Hard corruptions on court opinions. Prompting the model to verify support at the cited page (page-grounded prompting) raises recall on Hard examples by $10$--$28$~pp on court opinions and $7$--$36$~pp on briefs, at the cost of higher false positive rates.

\vspace{-0.18in}
\paragraph{Our Contributions.}
In this paper, we formalize \emph{legal citation support verification} as a task distinct from fabrication detection, and develop a controlled methodology for measuring it. We create a three-category taxonomy of citation roles---substantive, procedural, and secondary---in consultation with legal experts. We then operationalize it with an LLM classifier that we prompt-tune using expert annotations (\cref{sec:taxonomy}). Building on this taxonomy, we construct the test set over CLERC \citep{hou2024clerc} and BriefMe \citep{woo2025briefme} in which corruptions leverage the Bluebook structure to create difficulty-graded examples; the hardest examples hold the cited case fixed and change only the pinpoint page (\cref{sec:corruption-strategies}). Our evaluation of fourteen frontier model configurations across three families reveals that wrong-case substitutions are easy to detect while same-case pinpoint errors are hard. Extended reasoning and page-grounded prompting both reduce this gap without closing it (\cref{sec:prompt_intervention}).

We defer a fuller discussion of related work to \cref{appendix:related_work}.

\section{Task Definition and Citation Taxonomy}

\subsection{Task Definition}

Given a passage of legal text and the content of a cited document, the task is to determine whether the citation is \textit{on point}. That is, does the cited authority actually support the proposition for which it is cited? To indicate which citation in the passage is under review, we wrap it in \texttt{<CITE>}\,\texttt{</CITE>} tags before presenting the legal passage to the model. The prompt embeds two inputs. The first is the source paragraph with the target citation wrapped in the \texttt{<CITE>}\,\texttt{</CITE>} tags, and the second is the content of the cited document. When the citation includes a pinpoint page reference, the prompt embeds only the content of that page or page range; otherwise it receives the full document. We refer to these as \textit{pinpoint citations}. The model is asked to return a binary judgment (\texttt{YES}/\texttt{NO}) along with a rationale describing why it thinks the citation is or is not on point. The full prompt is provided in \cref{appendix:prompts}.

\subsection{Citation Taxonomy}
\label{sec:taxonomy}

As previously alluded to, legal documents cite authorities for many different reasons. But not all citations offer direct support to the proposition being made. Citations may also be used to trace the history of a case, or acknowledge a contrary position. Only a certain subset of citations make a support claim, so we formalize the task around these citations and filter out the rest. \Cref{appendix:legal_background} provides a short primer on the document types, Bluebook citation format, and the types of citations that commonly occur in legal writing. We worked with three legal analysts, all with prior litigation experience and familiarity with citation practice, to group citations into three types (\cref{tab:citation-type-stats} gives their distribution):

\begin{itemize}[leftmargin=*,itemsep=2pt]
    \item \textbf{Substantive citations:} These are citations that assert a legal rule and offer a case as the basis for that assertion. For instance, consider the following proposition, ``Due process requires notice reasonably calculated to apprise interested parties.'' \textit{Mullane v.\ Cent.\ Hanover Bank \& Tr.\ Co.}, 339 U.S. 306, 314 (1950). Here the citation supports the claim being made about ``due process''. These citations are therefore ones for which support errors are most consequential as they would invalidate the claim being made.
    \item \textbf{Procedural history citations:} These are citations that record what happened earlier in the history of the case. They trace the path the case took through the courts rather than supply the authority for a legal rule. For example, ``The district court granted summary judgment, \textit{aff'd}, 893 F.3d 1107 (8th Cir.\ 2018).''
    \item \textbf{Secondary citations:} Finally, these are the citations that provide additional background, comparison, or contrary authority, often introduced by Bluebook signals such as \textit{see}, \textit{cf.}, or \textit{but see}. For example, ``\textit{See also Chevron U.S.A., Inc.\ v.\ Nat.\ Res.\ Def.\ Council}, 467 U.S. 837, 842--43 (1984).''
\end{itemize}

Our experiments in this paper are geared around only the substantive citations.

\paragraph{Automated Classification.} To filter out non-substantive citations, we use a prompt-based LLM classifier. We evaluate this prompt on a held-out validation set annotated by the same legal experts who developed the taxonomy, achieving an F1 score of $0.83$ (\cref{appendix:dataset_stats}). Substantive citations dominate both datasets ($69$--$70\%$), confirming that legal documents primarily cite authority as direct support for propositions (see \cref{tab:citation-type-stats}). We restrict to case law citations, excluding statutes, regulations, and other secondary sources, which we leave for future work.

\section{Evaluation Design}
\label{sec:evaluation-design}

\subsection{Source Datasets}

We draw from two established legal datasets: CLERC \citep{hou2024clerc} (court opinions, $\sim 2{,}000$ citations) and BriefMe \citep{woo2025briefme} (legal briefs, $\sim 750$ citations). For each source text, we extract all embedded citations and obtain the full text of cited documents. The CLERC dataset includes the preceding paragraphs as context, whereas BriefMe does not. This difference affects which corruption levels we can construct, as described below.

\subsection{Corruption Strategies}
\label{sec:corruption-strategies}
Rule-based corruption of real examples is a standard way to generate synthetic negatives in NLP \citep{kryscinski-etal-2020-evaluating,10.24963/ijcai.2024/687,gekhman-etal-2023-trueteacher}. We adapt this idea to legal citation support verification, exploiting the structure of Bluebook citations to construct corruptions at three difficulty levels. We take existing citations from the source paragraphs as valid \textit{on-point} citations and apply corruption strategies as illustrated in \cref{fig:corruption-examples}. As the replacement citation grows more topically similar to the original, detection demands proposition-level verification.

We use controlled perturbations of real citations rather than LLM-generated corruptions to preserve the original legal prose and real citation contexts while changing only the support relation. Pilot experiments with LLM-based corruption showed that generated passages often altered text associated with citations other than the target citation or introduced unrelated defects. We discuss this more in \cref{sec:failed-corruption}.

Human validation (\cref{sec:human-validation}) confirms high agreement between annotators and our heuristic labels ($84$--$88\%$ across both datasets), supporting the claim that the corrupted citations are usually not on point. We also constructed small exploratory datasets based on case-to-case treatment relations, such as distinguishing and criticizing; results on these datasets are reported in \cref{appendix:relational_results}.

\paragraph{Random Document Corruption (Easy).}
We replace the citation with one from an entirely different legal document. This tests basic relevance detection: the replacement cited document shares almost no topical overlap with the original proposition.

\paragraph{Same-Document Corruption (Medium).}
We replace the citation with a different citation found in the preceding paragraphs of the same source document. Both citations appear in related legal contexts, but the replacement does not support the specific proposition. Since BriefMe does not include preceding paragraphs, we cannot construct this corruption for that dataset.

\paragraph{Wrong Page Number Corruption (Hard).}
We change only the pinpoint page number within the same case (e.g., ``500 U.S. at 105'' is changed to ``500 U.S. at 112''). The citation still references the same case but points to a different page, where a different legal issue unrelated to the original proposition might be discussed. This is the hardest setting because it preserves the most case-level overlap while breaking page-level support.

\subsection{Human Validation}
\label{sec:human-validation}
We next ask whether these heuristic corruptions actually produce examples that are not on point. We took $23$ citations per dataset with a mix of on-point and not-on-point examples and asked three legal analysts to label each citation as on point or not on point. They were not shown the correct labels, since the goal was to measure agreement with anticipated heuristic labels. Of these $23$, $10$ citations were labeled by all three annotators to measure inter-annotator agreement metrics. \Cref{tab:iaa-main} reports inter-annotator agreement.

\begin{table}[!t]
\centering
\small
\setlength{\tabcolsep}{8pt}
\renewcommand{\arraystretch}{1.08}
\begin{tabular}{@{}lrr@{}}
\softtoprule
\textbf{Dataset} & \textbf{3-way Agree} & \textbf{Gwet's AC1} \\
\midrule
CLERC & $90\%$ & $0.88$ \\
BriefMe & $80\%$ & $0.85$ \\
\bottomrule
\end{tabular}
\caption{Inter-annotator agreement on citation validity. AC1 remains high despite class imbalance.}
\label{tab:iaa-main}
\end{table}

We use Gwet's AC1 \citep{Gwet2008ComputingIR} rather than Fleiss' $\kappa$ \citep{Fleiss1971MeasuringNS} because $\kappa$ is unstable under class imbalance \citep{pradhan2025llmjudge, Brennan1981CoefficientKS} (for BriefMe, $\kappa$ drops to $-0.07$ despite $80\%$ three-way agreement; see \cref{tab:iaa-extended}). Agreement with our heuristic labels (``on point'' for original examples, ``not on point'' for corrupted examples) averaged $88\%$ for CLERC and $84\%$ for BriefMe (per-annotator rates in \cref{tab:label-agreement}). Overall, the corruptions reliably produce examples that annotators judge ``not on point.'' We further analyze the agreement metrics by each corruption type in \Cref{appendix:human_annotation}. Agreement remains high across all corruption types. This alleviates the concern that more difficult corruption types (e.g., wrong page) might produce more ambiguous examples. More details on the annotation process are provided in \cref{appendix:human_annotation}.

\begin{table}[!t]
\centering
\small
\setlength{\tabcolsep}{4pt}
\renewcommand{\arraystretch}{1.06}
\begin{tabular}{@{}lrrrr@{}}
\softtoprule
 & \multicolumn{3}{c}{\textbf{Recall (\%)}} & \textbf{FPR (\%)} \\
\cmidrule(lr){2-4}
\textbf{Model} & \textbf{Easy} & \textbf{Med.} & \textbf{Hard} & \\
\midrule
GPT-4.1 & $97.6$ & $73.6$ & $38.4$ & $11.2$ \\
GPT-4.1-mini & $94.8$ & $76.3$ & $42.0$ & $20.0$ \\
GPT-4o & $99.8$ & $91.6$ & $60.6$ & $34.9$ \\
GPT-4o-mini & $98.9$ & $84.8$ & $59.5$ & $43.7$ \\
GPT-5 & $99.6$ & $87.5$ & $55.5$ & $11.6$ \\
GPT-5-mini & $96.2$ & $74.6$ & $43.6$ & $9.8$ \\
GPT-5.4 & $99.6$ & $90.2$ & $48.5$ & $15.6$ \\
GPT-5.4 (reasoning) & $99.8$ & $90.3$ & $59.8$ & $13.1$ \\
\midrule
Claude Sonnet 4 & $99.2$ & $88.6$ & $47.9$ & $12.8$ \\
Claude Sonnet 4.6 & $99.8$ & $93.0$ & $44.3$ & $7.7$ \\
Claude Opus 4.6 & $99.2$ & $81.2$ & $36.5$ & $4.1$ \\
\midrule
Gemini 2.5 Flash & $98.9$ & $83.2$ & $59.2$ & $11.9$ \\
Gemini 2.5 Pro & $99.5$ & $84.4$ & $58.3$ & $11.9$ \\
Gemini 3.1 Pro & $99.7$ & $88.5$ & $56.3$ & $11.0$ \\
\bottomrule
\end{tabular}
\caption{CLERC results. Recall is the percentage of corrupted citations flagged; FPR is the percentage of valid citations incorrectly flagged.}
\label{tab:main-results}
\end{table}

\vspace{-0.1in}
\section{Experiments}
\subsection{Models}

We evaluate fourteen frontier model configurations across three families, covering thirteen base models: GPT-4o, GPT-4o-mini, GPT-4.1, GPT-4.1-mini, GPT-5, GPT-5-mini, and GPT-5.4 \citep{openai2024gpt4o,openai2025gpt5}; Claude Sonnet 4, Claude Sonnet 4.6, and Claude Opus 4.6 \citep{anthropic2025claude}; and Gemini 2.5 Flash, Gemini 2.5 Pro, and Gemini 3.1 Pro \citep{google2025gemini}. For GPT-5.4, we test two reasoning effort settings: \textit{none} (standard inference) and \textit{high} (chain-of-thought reasoning), yielding two evaluated configurations. We report results with two prompts: a baseline prompt for all main results, and a page-grounded prompt intervention (\cref{sec:prompt_intervention}) that adds explicit verification instructions. Both prompts are provided in \cref{appendix:prompts}.

\subsection{Evaluation Metrics}

We report \textbf{Recall} on corrupted citations (fraction of erroneous citations correctly flagged) and \textbf{False Positive Rate (FPR)} on valid citations (fraction of valid citations incorrectly flagged). A good guardrail requires high recall and low FPR.

\section{Results}
\label{sec:results}

\subsection{Main Results}
We see that models are nearly saturated on wrong-case substitutions, but same-case pinpoint mismatches remain difficult. \Cref{tab:main-results} reports results on CLERC (court opinions), and \cref{tab:briefme-results} reports results on BriefMe (legal briefs) across all models.

\begin{table}[!t]
\centering
\small
\setlength{\tabcolsep}{5.2pt}
\renewcommand{\arraystretch}{1.06}
\begin{tabular}{@{}lrrr@{}}
\softtoprule
 & \multicolumn{2}{c}{\textbf{Recall (\%)}} & \textbf{FPR (\%)} \\
\cmidrule(lr){2-3}
\textbf{Model} & \textbf{Easy} & \textbf{Hard} & \\
\midrule
GPT-4.1 & $96.2$ & $51.5$ & $16.1$ \\
GPT-4.1-mini & $93.0$ & $59.2$ & $19.5$ \\
GPT-4o & $99.4$ & $79.6$ & $50.8$ \\
GPT-4o-mini & $99.1$ & $71.8$ & $52.0$ \\
GPT-5 & $100.0$ & $77.0$ & $14.4$ \\
GPT-5-mini & $96.2$ & $59.0$ & $13.9$ \\
GPT-5.4 & $97.7$ & $64.0$ & $14.6$ \\
GPT-5.4 (reasoning) & $99.4$ & $82.0$ & $14.4$ \\
\midrule
Claude Sonnet 4 & $99.4$ & $68.6$ & $28.5$ \\
Claude Sonnet 4.6 & $99.4$ & $64.6$ & $18.6$ \\
Claude Opus 4.6 & $97.2$ & $54.0$ & $5.4$ \\
\midrule
Gemini 2.5 Flash & $97.7$ & $74.8$ & $12.0$ \\
Gemini 2.5 Pro & $99.1$ & $82.7$ & $16.3$ \\
Gemini 3.1 Pro & $99.6$ & $82.7$ & $11.9$ \\
\bottomrule
\end{tabular}
\caption{BriefMe results. Medium difficulty is not available for this dataset.}
\label{tab:briefme-results}
\end{table}

\paragraph{Recall drops sharply on same-case pinpoint corruptions.}
Recall on Easy examples is uniformly high. It ranges from $94.8$--$99.8\%$ on CLERC and $93.0$--$100.0\%$ on BriefMe. On CLERC, recall on Medium examples falls to $73.6$--$93.0\%$. Recall on Hard examples falls further to $36.5$--$60.6\%$. On BriefMe, recall on Hard examples ranges from $51.5$--$82.7\%$. This matches our expectation that pinpoint corruptions would be harder to detect than the other corruption types (\cref{fig:corruption-examples,sec:corruption-strategies}). Further analysis in \Cref{sec:error-analysis} shows that models often treat topical overlap as sufficient support.

\paragraph{Larger models improve, but not predictably.}
Within the OpenAI family of models, we see a better recall on Hard examples with GPT-5 than with GPT-4.1 on both CLERC ($55.5\%$ vs. $38.4\%$) and BriefMe ($77.0\%$ vs. $51.5\%$) testsets. This is consistent with the view that GPT-5 is a more capable model, but the metrics do not extrapolate monotonically with model capability. GPT-5.4 without reasoning trails behind GPT-5 on both testsets. Claude Opus 4.6, which is Anthropic's more capable model, also trails both Sonnet variants on Hard-examples, despite being the nominally more capable model.

\paragraph{Higher recall comes with more false positives.}
On the CLERC testset, GPT-4o attains the highest baseline recall on Hard examples at $60.6\%$. Its FPR is $34.9\%$, so it incorrectly flags roughly one in three valid citations. Claude Opus 4.6, on the other hand, attains the lowest FPR ($4.1\%$ on CLERC and $5.4\%$ on BriefMe). However, it also has the lowest recall on Hard examples on both testsets, at $36.5\%$ and $54.0\%$. The same tradeoff is visible in the BriefMe testset. The highest recall on Hard examples is $82.7\%$, achieved by Gemini 2.5 Pro and Gemini 3.1 Pro, but neither model matches the lowest-FPR setting. In \Cref{sec:prompt_intervention}, we return to analyzing this tradeoff further under a stricter prompting setting.

\paragraph{Recall is higher on briefs than on opinions.}
Across all models we study, recall on Hard examples is higher on BriefMe than on CLERC, by $12$--$26$~pp. We observe the same pattern when reasoning is turned on. For GPT-5.4, high reasoning effort adds $18.0$~pp on BriefMe and $11.3$~pp on CLERC. We return to possible reasons for this document-type gap in \cref{sec:discussion}. \Cref{appendix:legal_background:documents} summarizes some fundamental difference between court opinions and briefs.

\paragraph{Reasoning helps, but the gap remains.}
Adding high reasoning effort to GPT-5.4 increases recall on Hard examples from $48.5\%$ to $59.8\%$ on CLERC and from $64.0\%$ to $82.0\%$ on BriefMe. It also lowers FPR slightly, from $15.6\%$ to $13.1\%$ on CLERC and from $14.6\%$ to $14.4\%$ on BriefMe. Reasoning does lead to an overall improvement, but it leaves the hardest setting far from solved: on CLERC, the model still misses about $40\%$ of pinpoint mismatches. In later experiments in \Cref{sec:prompt_intervention}, we show that stronger page-grounded prompting improves recall further, but it also raises the false positive rate.

\begin{takeawaybox}
Models are near-saturated on Easy examples, but recall on Hard examples falls to $36.5$--$60.6\%$ on court opinions and $51.5$--$82.7\%$ on briefs. No model achieves both low FPR and high recall on Hard examples.
\end{takeawaybox}

\section{Diagnosing the Failure Mode}

\subsection{Failure Analysis}
\label{sec:error-analysis}

Here, we discuss the three recurring error patterns that we observed in the course of our failure analysis: invented support rationales, topical matching, and failure to check quoted text. To make these patterns concrete, we examine GPT-5 false negatives, i.e., corrupted citations the model incorrectly accepts as ``on-point''. Consider the following Hard example for which the correct page is $364$ but we replaced it with page $371$.

\begin{tcolorbox}[
    breakable,
    colback=gray!3,
    colframe=gray!50,
    boxrule=0.5pt,
    arc=2pt,
    left=3mm,
    right=3mm,
    top=2mm,
    bottom=2mm]
\small
\textit{``The bespeaks caution doctrine `merely reflects the unremarkable proposition that statements must be analyzed in context.'\,''} Rubinstein v.\ Collins, 20 F.3d 160, 169 (5th Cir.\ 1994) (citing In re Donald J.\ Trump Casino Sec.\ Litig., 7 F.3d 357, \textcolor{red}{371} (3d Cir.\ 1993)).

\vspace{0.3em}
\textbf{Cited page content (p.\ $371$):} ``The `Special Considerations' section also detailed the high level of competition for customers the completed Taj Mahal would face once opened to the public: Competition in the Atlantic City casino/hotel market is intense\ldots''
\end{tcolorbox}

\noindent The quoted language about ``statements must be analyzed in context'' does not appear on page $371$. That page instead discusses casino competition in Atlantic City. Yet GPT-5 accepts the citation. It asserts that the target content ``explicitly says, `we must consider an alleged misrepresentation within the context in which it was communicated'\,'', fabricating a paraphrase that does not appear on the cited page. This is not an isolated case. \Cref{tab:error-examples} in \cref{appendix:error_analysis} shows additional examples of the same failure pattern.

The error analysis exercise reveals the following findings.

\paragraph{Invented support rationales.} Roughly two-thirds of these missed corruptions have a rationale that says the cited page ``expressly states'' or ``explicitly says'' text that is not there. The absence of this language on the cited page should lead the model to reject the citation. Instead, it claims that the cited page explicitly supports or states the proposition. This confident wording could make this fabricated rationale seem credible to a lawyer reviewing it.

\paragraph{Topical matching instead of propositional verification.} In our Hard corruptions we swap pages within the same case. This results in target content that remains topically related to the source passage. It is often the same dispute, and often the same statute or doctrine. The model treats this overlap as sufficient, rather than checking whether the cited page states the precise proposition being made. Easy corruptions replace the case entirely, so the topical mismatch is obvious and easy to detect.

\paragraph{Checking verbatim quotes.} Most Hard examples contain verbatim quotes attributed to the cited authority. Among GPT-5 false negatives on these examples, the quoted text does not appear anywhere on the cited page $92\%$ of the time. This observation motivates a simple prompt change: explicitly asking the model to verify the presence of quoted text, which we explore next.

\subsection{Page-Grounded Prompt Intervention}
\label{sec:prompt_intervention}
We designed a modified prompt (\cref{appendix:v2prompt}) targeting the failure patterns above. To address the \emph{verbatim quoted text} failure, it asks the model to check if verbatim-quoted language appears in the target content. To address the \emph{topical matching problem}, it asks if the specific target content supports the claim, rather than accepting the citation for topical relevance. To address the \emph{invented support rationales} failure mode, it instructs the model not to say that the target content ``expressly states'' something unless those words, or a close paraphrase, appear there. We also added a negative in-context example showing a topically related but wrong page from the same case. All other prompt elements remain identical.

\begin{table}[!t]
\centering
\small
\setlength{\tabcolsep}{3.7pt}
\renewcommand{\arraystretch}{1.06}
\begin{tabular}{@{}lrrrr@{}}
\softtoprule
 & \multicolumn{2}{c}{\textbf{Hard Recall (\%)}} & \multicolumn{2}{c}{\textbf{FPR (\%)}} \\
\cmidrule(lr){2-3}\cmidrule(lr){4-5}
\textbf{Model} & \textbf{Base} & \textbf{+Gnd.} & \textbf{Base} & \textbf{+Gnd.} \\
\midrule
GPT-4.1 & $38.4$ & $66.2$\pmark{+27.8} & $11.2$ & $22.7$\nmark{+11.5} \\
GPT-4.1-mini & $42.0$ & $58.7$\pmark{+16.7} & $20.0$ & $28.0$\nmark{+8.0} \\
GPT-4o & $60.6$ & $77.5$\pmark{+16.9} & $34.9$ & $49.3$\nmark{+14.4} \\
GPT-4o-mini & $59.5$ & $82.6$\pmark{+23.1} & $43.7$ & $64.2$\nmark{+20.5} \\
GPT-5 & $55.5$ & $68.8$\pmark{+13.3} & $11.6$ & $16.9$\nmark{+5.3} \\
GPT-5-mini & $43.6$ & $63.4$\pmark{+19.8} & $9.8$ & $16.4$\nmark{+6.6} \\
GPT-5.4 & $48.5$ & $66.3$\pmark{+17.8} & $15.6$ & $22.6$\nmark{+7.0} \\
GPT-5.4 (reasoning) & $59.8$ & $73.2$\pmark{+13.4} & $13.1$ & $19.5$\nmark{+6.4} \\
\midrule
Claude Sonnet 4 & $47.9$ & $57.6$\pmark{+9.7} & $12.8$ & $18.8$\nmark{+6.0} \\
Claude Sonnet 4.6 & $44.3$ & $62.0$\pmark{+17.7} & $7.7$ & $14.0$\nmark{+6.3} \\
Claude Opus 4.6 & $36.5$ & $51.5$\pmark{+15.0} & $4.1$ & $5.1$\nmark{+1.0} \\
\midrule
Gemini 2.5 Flash & $59.2$ & $77.9$\pmark{+18.7} & $11.9$ & $36.4$\nmark{+24.5} \\
Gemini 2.5 Pro & $58.3$ & $72.9$\pmark{+14.6} & $11.9$ & $19.5$\nmark{+7.6} \\
Gemini 3.1 Pro & $56.3$ & $67.7$\pmark{+11.4} & $11.0$ & $13.7$\nmark{+2.7} \\
\bottomrule
\end{tabular}
\caption{Effect of the page-grounded prompt on CLERC. Green superscripts mark recall gains; red superscripts mark FPR increases.}
\label{tab:ablation-clerc}
\end{table}

\begin{table}[!t]
\centering
\small
\setlength{\tabcolsep}{3.7pt}
\renewcommand{\arraystretch}{1.06}
\begin{tabular}{@{}lrrrr@{}}
\softtoprule
 & \multicolumn{2}{c}{\textbf{Hard Recall (\%)}} & \multicolumn{2}{c}{\textbf{FPR (\%)}} \\
\cmidrule(lr){2-3}\cmidrule(lr){4-5}
\textbf{Model} & \textbf{Base} & \textbf{+Gnd.} & \textbf{Base} & \textbf{+Gnd.} \\
\midrule
GPT-4.1 & $51.5$ & $87.0$\pmark{+35.5} & $16.1$ & $30.8$\nmark{+14.7} \\
GPT-4.1-mini & $59.2$ & $76.0$\pmark{+16.8} & $19.5$ & $30.6$\nmark{+11.1} \\
GPT-4o & $79.6$ & $89.0$\pmark{+9.4} & $50.8$ & $59.4$\nmark{+8.6} \\
GPT-4o-mini & $71.8$ & $93.0$\pmark{+21.2} & $52.0$ & $71.7$\nmark{+19.7} \\
GPT-5 & $77.0$ & $91.0$\pmark{+14.0} & $14.4$ & $22.5$\nmark{+8.1} \\
GPT-5-mini & $59.0$ & $89.0$\pmark{+30.0} & $13.9$ & $26.7$\nmark{+12.8} \\
GPT-5.4 & $64.0$ & $91.0$\pmark{+27.0} & $14.6$ & $24.2$\nmark{+9.6} \\
GPT-5.4 (reasoning) & $82.0$ & $89.0$\pmark{+7.0} & $14.4$ & $21.1$\nmark{+6.7} \\
\midrule
Claude Sonnet 4 & $68.6$ & $85.0$\pmark{+16.4} & $28.5$ & $38.5$\nmark{+10.0} \\
Claude Sonnet 4.6 & $64.6$ & $90.7$\pmark{+26.1} & $18.6$ & $24.4$\nmark{+5.8} \\
Claude Opus 4.6 & $54.0$ & $77.3$\pmark{+23.3} & $5.4$ & $7.1$\nmark{+1.7} \\
\midrule
Gemini 2.5 Flash & $74.8$ & $91.0$\pmark{+16.2} & $12.0$ & $36.7$\nmark{+24.7} \\
Gemini 2.5 Pro & $82.7$ & $90.2$\pmark{+7.5} & $16.3$ & $28.6$\nmark{+12.3} \\
Gemini 3.1 Pro & $82.7$ & $89.2$\pmark{+6.5} & $11.9$ & $19.0$\nmark{+7.1} \\
\bottomrule
\end{tabular}
\caption{Effect of page-grounded prompt on BriefMe (legal briefs). Improvements are larger than on CLERC (court opinions).}
\label{tab:ablation-briefme}
\end{table}

\Cref{tab:ablation-clerc,tab:ablation-briefme} show the effect of the page-grounded prompt. We comment on the observations below.

\paragraph{Page-grounded prompting improves recall on Hard examples.} Across all models, we see an improvement: $9.7$--$27.8$ pp on CLERC and $6.5$--$35.5$ pp on BriefMe. The gains are concentrated on Hard examples. Recall on Easy examples, already saturated above $95\%$ for most models, improves only by $0.7$ pp on CLERC. The size of the gain confirms that the baseline prompt was not eliciting page-level verification, and it is not merely attributable to string matching: on GPT-5, recall improves by $+14.1$ pp on cases with verbatim quotes versus $+10.5$ pp on cases without.

\paragraph{Page-grounded prompting also raises false positive rates.} This is the same tradeoff we noted earlier in the baseline results in the last section. Across the two datasets, the false positive rate rises by $1.0$--$24.7$ pp. The additional instructions appended to the prompt make the models more skeptical of \textit{all} citations, not just the corrupted ones. The largest FPR increases occur for GPT-4o-mini and Gemini 2.5 Flash, whereas Claude Opus 4.6 changes the least.

\begin{takeawaybox}
The recall improvement as a result of page-grounded prompting is accompanied by a corresponding increase in false positive rates. The intervention makes the models more skeptical overall, not more selective.
\end{takeawaybox}

\section{Discussion}
\label{sec:discussion}
We organize the discussion around the three research questions posed in \cref{sec:introduction}, then discuss broader implications.

\paragraph{RQ1: Detection by Difficulty.}
Models perform well at detecting Easy errors. But performance on Hard examples remains much lower ($37$--$61\%$ on CLERC and $52$--$83\%$ on BriefMe). This drop of $16$--$63$~pp from Easy to Hard recall across both corpora suggests that current LLMs struggle with proposition-level verification.

\paragraph{RQ2: Model Scale and Reasoning.}
Within the OpenAI family, GPT-5 outperforms GPT-4.1 on Hard examples ($55.5\%$ vs.\ $38.4\%$ on CLERC and $77.0\%$ vs.\ $51.5\%$ on BriefMe), but the improvement is modest relative to the overall gap. The gains also do not extrapolate monotonically for more capable models: GPT-5.4 without reasoning trails GPT-5 on both datasets, and Claude Opus 4.6 trails both Sonnet variants despite being the nominally more capable model. We also observe a positive effect of reasoning: high reasoning effort helps GPT-5.4, raising it to $59.8\%$ recall with $13.1\%$ FPR on CLERC and $82.0\%$ recall with $14.4\%$ FPR on BriefMe. Yet even in this setting, it still misses $40\%$ of pinpoint mismatches on court opinions. We therefore conclude that differentiating between valid and invalid citations at the proposition level remains a challenging problem for current LLMs, even with improved reasoning.

\paragraph{RQ3: Document Type.}
Finally, we observe that recall on Hard examples is $12$--$26$~pp higher on briefs than on court opinions across all models, and extended reasoning shows the largest improvement on briefs ($+18$~pp for GPT-5.4). One possible explanation is that briefs are advocacy documents and may often state narrower propositions than court opinions, so a wrong page may create a clearer mismatch. We did not directly measure proposition specificity, so we treat this explanation as a hypothesis rather than a tested explanation, and leave it to future work.

\paragraph{The Failure Mode and Its Limits.}
Our experiments confirm that models rely on topical coherence more readily than on proposition-level support. A page-number mismatch creates a setting where the cited content remains doctrinally and topically related to the proposition, but does not actually support it.

This resembles a broader tendency of LLMs to preserve coarse semantic categories while blurring finer-grained distinctions that humans retain \citep{shani2025tokens}. It also echoes the findings from cognitive psychology that ease of processing can serve as a cue to truth in human judgment, though we cite this literature only as an analogy rather than as a direct claim for legal citation verification \citep{reber2010epistemic}.

Referring back to the error analysis in \cref{sec:error-analysis}, we saw that in roughly two-thirds of missed pinpoint mismatches, the model claims that the cited page ``expressly states'' text that supports the proposition, even though that text does not appear there. Among missed Hard corruptions containing verbatim quotes, the quoted language is absent from the cited page $92\%$ of the time.

The page-grounded prompt (\cref{sec:prompt_intervention}) improved recall on Hard examples. This indicates that the baseline prompt was not eliciting proposition-level verification. However, these gains come with increased false positives. This tendency undermines the reliability of prompt-only citation support verification, highlighting the need for mechanisms that verify whether the cited page supports the specific proposition. The intervention is more of a diagnostic rather than a corrective measure. Models can recover some missed pinpoint mismatches when pushed toward proposition-level checking, but they do so partly by treating valid citations as suspicious.

\paragraph{Why Legal Citation Verification Is Structurally Different.}

In general fact verification benchmarks such as FEVER \citep{thorne-etal-2018-fever}, claims are short and evidence is localized to the sentence or paragraph level; in legal citation verification, claims may be longer, often including several subclauses or spanning multiple sentences, and the relevant authority may be much larger, spanning the entire case or a large page range. Checking for citation support is not clear-cut entailment; it may depend on legal analogy, precedent, or the scope of a holding. An NLI model may treat topical overlap as support, but citation checking asks whether the cited page supports the specific proposition. This suggests that support checking may need two steps: first locating the relevant content in the cited authority, and then judging whether that content supports the asserted proposition \citep{yuan2026citeaudit}.

\paragraph{Limits of LLM-Generated Corruptions.}\label{sec:failed-corruption}
We initially explored LLM-based corruption by asking a model to subtly modify the source passage while keeping the citation unchanged. Human evaluation showed that these generations often altered text unrelated to the proposition for that citation, producing invalid test cases. This negative result is itself informative. If an LLM cannot reliably isolate the citation-relevant claim when generating corruptions, it is unsurprising that it also struggles to isolate that claim during verification.

\paragraph{Implications for Legal Practice.}
In \textit{Mata v.\ Avianca}, the court sanctioned attorneys for submitting fabricated cases. A case database lookup can catch those errors. The more insidious risk is a citation to a real case, correctly cited with reporter and pinpoint information, that does not support the claimed proposition. This is the scenario our controlled experiments simulate.

Our results should not be read to mean that frontier models lack utility for citation verification. In practice, lawyers spend time reviewing model output or human-drafted legal work before accepting it. Citation support verification tools that clearly separate easy cases from difficult proposition-level verification can still be useful in reducing review time.
Our results suggest that shallow review, limited to checking case existence or accepting the model's rationale at face value, may be insufficient to catch the errors we identify.

Recent cases also suggest that this burden applies not only to lawyers' own filings, but also to an opponent's submissions.
In \textit{Noland v.\ Land of the Free, L.P.}, the California Court of Appeal declined to award sanctions payable to opposing counsel because respondents did not alert the court to the fabricated citations and appeared to have become aware of the problem only after the court issued an order to show cause \citep{nolandcase}. Read together with \textit{Mata}, \textit{Noland} raises the possibility that verification obligations may extend to opposing parties' filings as well, though this area of law is still developing. For bar associations and courts developing AI use guidelines, our results raise the question of whether AI-assisted citation checking that operates below proposition-level verification provides adequate safeguards.

Prior work on AI-assisted lawyering suggests that AI tools can improve productivity when lawyers use them judiciously and under supervision \citep{choi2024lawyering}. Our findings identify a boundary condition for that view: tools that verify only the existence of a cited authority would miss the misrepresentation errors we study, where a real citation fails to support the asserted proposition.

\vspace{-0.1in}
\section{Limitations}
\vspace{-0.1in}
As outlined earlier, our study has several scope limitations. We focus on substantive citations in U.S. federal court opinions and briefs. We treat citation support as a binary decision, either on point or not on point. In practice, attorneys make a more nuanced assessment and evaluate citations on a host of criteria that include binding versus persuasive authority, degree of factual analogy, and whether subsequent decisions have upheld or overruled the cited authority. Our evaluation targets the clearest case---whether the cited page supports the proposition made in the source text---it does not capture these finer-grained distinctions.

Our measurements also involve several caveats. In our human validation study, annotators agreed in $84$--$88\%$ of cases that the corruptions produce not-on-point citations.
Some corrupted citations may nevertheless remain on point by chance. If a model accepts such a citation, our evaluation counts it as a missed error, so recall on corrupted examples may be understated.
The difficulty labels associated with our corruption strategies are also coarse approximations. They do not cover all plausible forms of citation misrepresentation, including relational substitutions between closely connected cases such as authorities that cite, distinguish, or criticize one another. We have some preliminary experiments in \cref{appendix:relational_results} that suggest these may be even harder for models to detect than the wrong-pinpoint errors we study here, but a more systematic evaluation of this important failure mode is left for future work.

Finally, we do not compare model-assisted legal work against human legal-review workflows or measure end-to-end auditing costs. In practice, lawyers may use LLMs under supervision rather than filing raw model output. False positives increase reviewer burden, while false negatives create false reassurance; accordingly, our per-citation verification results do not fully capture the utility of LLMs for auditing citations in realistic supervision workflows.
\vspace{-0.1in}
\section{Conclusion}
\vspace{-0.1in}
Database lookup largely solves the fake citation problem. The harder scenario is a real case citation whose cited content does not support the proposition in the source text. That is the support-verification problem we study in this paper.

Across fourteen model configurations from three model families and two document types, recall on Easy examples is near-saturated while recall on Hard examples falls by $16$--$63$~pp. Using more capable models or extended reasoning does not close this gap. GPT-5.4 with high reasoning effort still misses $40\%$ of pinpoint mismatches on court opinions. Page-grounded prompting improves recall on Hard examples, but those gains come with higher false positive rates. Current models can be made more skeptical, but not selectively skeptical enough to distinguish valid from invalid citations at the proposition level.

From a legal practice perspective, checking AI-generated legal text cannot stop at verifying that cited cases exist. It must also test for proposition-level support. Our results show that current models often recognize the right legal topic without verifying support for the cited proposition, and this latter capability remains difficult for current models. Closing that gap may require more than prompt tuning, larger models, or extended reasoning.

\vspace{-0.7em}
\section*{Acknowledgments}
\vspace{-0.5em}
The author thanks Satyajit Dixit for a thorough review; Robert Kingan, Madhavan Seshadri, and Fulya Erdinc for draft review; Leo Chen and Arnav Mantro for engineering implementation, whose careful work made the empirical pipeline substantially more reliable; and Pushpit Saxena for helpful discussions. This paper benefited from a team effort and from the judgment and generosity of many colleagues. The author also thanks the annotators and others from whom he learned through discussions.

\section*{Reproducibility Statement}

We describe our evaluation methodology in detail in \cref{sec:evaluation-design,appendix:dataset_format}. Prompt templates are provided in \cref{appendix:prompts}. Both source corpora (CLERC, BriefMe) are publicly available.

\section*{Ethics and Impact Statement}

This work aims to improve legal AI safety by showing where citation verification fails. All data come from public legal corpora (court opinions and briefs). Our findings could help adversaries craft harder-to-detect errors. We think the benefit of exposing these limitations outweighs this risk. Human annotators were legal analysts with prior litigation experience, compensated at market rates.

\bibliography{custom}
\bibliographystyle{icml2026}

\appendix
\clearpage
\onecolumn
\appendix


\begin{adjustwidth}{5em}{5em}

\section*{Appendix Contents}

\vspace{0.8em}

{%
\hypersetup{linkcolor=black}%
\makeatletter
\newcommand{\apptocentry}[3]{%
  \noindent{\color{blue!70!black}\textbf{#1\hspace{1em}#2}}%
  {\color{black}\leaders\hbox to 0.8em{\hss.\hss}\hfill\mbox{\textbf{\pageref{#3}}}}\par\vspace{0.4em}}
\newcommand{\apptocsub}[3]{%
  \noindent\hspace*{1.5em}{\color{blue!70!black}\makebox[2.3em][l]{#1}#2}%
  {\color{black}\leaders\hbox to 0.8em{\hss.\hss}\hfill\mbox{\pageref{#3}}}\par\vspace{0.3em}}
\makeatother
\apptocentry{A}{Related Work}{appendix:related_work}

\vspace{0.3em}
\apptocentry{B}{Prompts}{appendix:prompts}
\apptocsub{B.1}{Citation Verification Prompt (Baseline)}{appendix:prompts:baseline}
\apptocsub{B.2}{Page-Grounded Verification Prompt}{appendix:v2prompt}

\vspace{0.3em}
\apptocentry{C}{Dataset Statistics}{appendix:dataset_stats}

\vspace{0.3em}
\apptocentry{D}{Human Annotation Details}{appendix:human_annotation}

\vspace{0.3em}
\apptocentry{E}{Error Analysis}{appendix:error_analysis}

\vspace{0.3em}
\apptocentry{F}{Evaluation Data Format}{appendix:dataset_format}
\apptocsub{F.1}{Processing Beyond Source Datasets}{appendix:contributions}
\apptocsub{F.2}{Schema Definition}{appendix:schema}
\apptocsub{F.3}{Example: BriefMe Easy Corruption}{appendix:example_briefme}
\apptocsub{F.4}{Example: CLERC Hard Corruption}{appendix:example_clerc}

\vspace{0.3em}
\apptocentry{G}{Full Prompt Sensitivity Results}{appendix:ablation_full}
\apptocsub{G.1}{CLERC (Court Opinions)}{appendix:ablation_clerc}
\apptocsub{G.2}{BriefMe (Legal Briefs)}{appendix:ablation_briefme}

\vspace{0.3em}
\apptocentry{H}{Diagnostic Results on Distinguished and Criticized Authorities}{appendix:relational_results}

\vspace{0.3em}
\apptocentry{I}{Primer on U.S. Legal Citation and Document Types}{appendix:legal_background}
\apptocsub{I.1}{Document Types}{appendix:legal_background:documents}
\apptocsub{I.2}{Anatomy of a Case Citation}{appendix:legal_background:anatomy}
\apptocsub{I.3}{Common Citation Patterns}{appendix:legal_background:patterns}
\apptocsub{I.4}{Other Terms Used in This Paper}{appendix:legal_background:terms}
}
\end{adjustwidth}

\newpage

\section{Related Work}
\label{appendix:related_work}

\paragraph{Legal Citation and Retrieval.} Prior legal NLP work has addressed adjacent parts of the citation problem, including finding relevant authority and formatting citations correctly. \citet{Mahari_Stammbach_Ash_Pentland_2024} frame legal passage retrieval as a core challenge and report only $59\%$ recall for their best models on LePaRD. Our Hard cases ask whether, given a passage and a citation, a model can verify that the citation supports the passage. Our recall on Hard examples ($37$--$61\%$) is close to their retrieval ceiling. This comparison suggests that proposition-level support verification remains difficult even when the cited authority is already identified. A separate line of work studies citation form. \citet{dahl2025bluebook} asks whether LLMs can produce correctly formatted legal citations. We move from citation form to citation support by asking whether a correctly formatted citation is substantively supported. CLERC \citep{hou2024clerc} and BriefMe \citep{woo2025briefme} provide the citation-level datasets for our study.

\paragraph{Legal Hallucination and Faithfulness.} This support question is central to legal hallucination work. \citet{Dahl_2024} distinguish ``fabricated'' from ``misattributed'' hallucinations. We stress-test the latter category, which they argue is harder to detect. That distinction matters for current legal research tools because fabricated authorities and misattributed support pose different verification problems. \citet{magesh2025hallucination} ask whether AI legal research tools are ``hallucination-free.'' Concurrent work by \citet{liu2026checks} introduces a legal citation hallucination benchmark and finds that even agentic checkers remain weakest on incorrect pincites and content misrepresentations. Our results help explain why that claim remains too broad. Current systems may catch fabrications, which a database lookup can often expose, yet still miss misrepresentations that require proposition-level verification. This also connects to broader concerns about unstable legal reasoning. \citet{purushothama2025bench} show that LLM legal interpretation is unstable and weakly calibrated to human judgments. Our finding is complementary because models struggle even when the task is verification rather than open-ended interpretation.

\paragraph{Citation Evaluation in NLP.} The closest methodological analogues come from citation evaluation outside legal NLP.
\citet{gao-etal-2023-enabling} introduce the ALCE benchmark for evaluating LLM-generated citations and find that even the best models lack complete citation support $50\%$ of the time.
This motivates evaluation beyond citation presence.
\citet{xu2025citeeval} argue that NLI-based citation evaluation treats citations as binary (supports/does not support) when citation support is more nuanced.
Related work shows why that nuance matters.
\citet{Sarol_Ming_Radhakrishna_Schneider_Kilicoglu_2024} study quotation errors in biomedical publications, while \citet{wojtasik2025citeverifier} show in RAG settings that verifier behavior depends on whether evaluation requires complete or partial citation support.
Our results support these concerns in the legal domain because frontier LLMs fall into the same topical-matching failure mode seen in NLI-style evaluation.
\citet{press2024citeme} test whether LLMs can identify the paper being cited given a text excerpt.
Frontier models achieve $4$--$19\%$ accuracy, compared with $70\%$ for humans, and the dominant error mode is selecting a topically similar paper rather than the correct one.
\citet{choi2026citeguard} extend this line with a retrieval-aware citation attribution agent, showing that additional claim context and within-paper search improve scientific citation attribution.
We observe a related pattern in legal verification, where models accept citations from the right case but the wrong page.
A parallel line of work studies citation auditing in scientific writing as an end-to-end verification problem \citep{yuan2026citeaudit}.
At the venue level, \citet{sakai2026hallucitation} identify nearly $300$ papers with at least one hallucinated reference across ACL, NAACL, and EMNLP 2024--2025, while \citet{xu2026ghostcite} report invalid citations in $604$ of $56{,}381$ papers across major AI/ML and security venues.
\citet{rao2026detecting} measure URL-level citation fabrication across $221{,}000$ URLs, finding $3$--$13\%$ hallucination rates even with retrieval augmentation.
CiteAudit focuses on scientific references and hallucinated or mismatched citations \citep{yuan2026citeaudit}.
Our study assumes the legal authority is real and asks whether the cited page supports the claimed proposition.

\paragraph{Hallucination Detection.} More broadly, our results qualify work on LLM factuality and verification. A growing body of work evaluates LLM factuality, from atomic-fact decomposition \citep{min-etal-2023-factscore} and claim verification against source documents \citep{wadden-etal-2020-fact} to comprehensive surveys of hallucination types and detection methods \citep{huang2025survey}. \citet{guan2024language} show that LLMs can serve as effective fact verifiers even when they are unreliable generators. Our work tests this verifier capability in legal citation, where topical similarity does not imply semantic support. The general-domain finding that LLMs excel at verification does not transfer here. When the cited authority discusses the right legal topic but makes a different point, frontier models accept the citation $40$--$63\%$ of the time.

\newpage

\section{Prompts}
\label{appendix:prompts}

\subsection{Citation Verification Prompt (Baseline)}
\label{appendix:prompts:baseline}

\begin{nolinenumbers}
\begin{tcolorbox}[width=5.5in, center, colback=blue!5!white, colframe=gray!50!black, title=Citation Verification Prompt, breakable, enhanced jigsaw]
\begin{lstlisting}[style=prompt]
Evaluate whether a given citation supports a legal claim made within a source paragraph, using the IRAC (Issue, Rule, Analysis, Conclusion) framework.

Instructions:
- You will receive two inputs:
  1. src_paragraph: The source paragraph with the citation enclosed in <CITE> </CITE> tags.
  2. tgt_content: Content from the cited document or specific pages indicated by the pinpoint citation.
- Use the IRAC framework:
  - Issue: Clearly state if the cited document addresses the specific legal issue raised by the src_paragraph.
  - Rule: Identify the legal rule or principle in the cited content relevant to the issue.
  - Analysis: Analyze whether the cited content explicitly or implicitly supports, contradicts, or is irrelevant to the specific citation in the src_paragraph.
  - Conclusion: State definitively if the citation is "on point" (YES) or not "on point" (NO).
- Provide a brief, clear rationale for your conclusion.

Positive Example:
src_paragraph: "[C]reditors are entitled to recover attorney's fees in bankruptcy [] if they have a contractual right to [such fees] valid under state law. <CITE> 456 F.3d 668 </CITE>"
tgt_content: "...creditors are entitled to recover attorney's fees in bankruptcy claims if they have a contractual right valid under state law..."
Output: { "on_point": "YES", "rationale": "The cited document explicitly states creditors can recover attorney's fees in bankruptcy if supported by valid contractual rights under state law." }

Negative Example:
src_paragraph: "Courts consistently refuse attorney's fees in bankruptcy where no explicit contractual provision exists. <CITE> 549 U.S. 443 </CITE>, 127 S. Ct. 1199, 167 L. Ed. 2d 178 (2007)"
tgt_content: "The Court determined that, absent specific statutory guidance, fees should be considered on a case-by-case basis, evaluating fairness and the equities involved."
Output: { "on_point": "NO", "rationale": "The cited document discusses a case-by-case analysis rather than explicitly supporting refusal of attorney's fees without contractual provisions." }

Additional Guidance:
- Consider the explicitness, context, and jurisdictional relevance of the citation.
- Ambiguity or indirect references generally suggest the citation is not on point.
- The rationale should succinctly justify your conclusion based on the content provided.

Now generate your output for the following input. Provide JSON-formatted response only:
INPUT:
Statement: $src_paragraph
Target Content: $tgt_content
OUTPUT:
\end{lstlisting}
\end{tcolorbox}
\end{nolinenumbers}

\subsection{Page-Grounded Verification Prompt}
\label{appendix:v2prompt}

This prompt modifies the baseline prompt above with three additional verification steps. All other elements of the prompt remain identical. This was used in \cref{sec:prompt_intervention} for the page-grounded intervention.

\begin{nolinenumbers}
\begin{tcolorbox}[width=5.5in, center, colback=blue!5!white, colframe=gray!50!black, title=Page-Grounded Verification Prompt, breakable, enhanced jigsaw]
\begin{lstlisting}[style=prompt]
[Instructions identical to baseline prompt above.]

[Baseline checklist omitted for brevity.]

Critical Verification Steps:

Before concluding that a citation is on point, you MUST perform these checks:

1. Verbatim quote check: If the src_paragraph quotes specific language, then verify that the quoted language actually appears in the tgt_content. If the quoted text is not present, the citation is NOT on point regardless of topical relevance.

2. Page-level verification: Do not accept a citation merely because the cited case is about the right legal topic. The tgt_content represents the specific page(s) cited. Verify that THIS specific content--not the case generally--supports the specific proposition.

3. Distinguish topic from support: A citation is on point only if the tgt_content supports the SPECIFIC claim being made, not merely if it discusses the same area of law.

Positive Example: [Same as baseline prompt above.]

Negative Example (topically related but wrong page):
src_paragraph: "The court held that `it is the duty of courts to harmonize jury answers where possible.' <CITE> 134 F.3d 1458, 1467 </CITE>"
tgt_content: "A fair and reasonable reading of the jury's verdict is that the jury chose to credit some or all of DMSI's business justifications..."
Output: { "on_point": "NO", "rationale": "The src_paragraph quotes specific language about courts' duty to harmonize jury answers. This quoted language does not appear in the tgt_content...", "confidence": 0.92 }

Negative Example and Additional Guidance: [Same as baseline, plus:]
- If you claim the tgt_content "expressly states" or "explicitly says" something, verify that those words or their close paraphrase actually appear in the tgt_content. Do not assert the presence of text you cannot locate.
\end{lstlisting}
\end{tcolorbox}
\end{nolinenumbers}

\newpage
\section{Dataset Statistics}
\label{appendix:dataset_stats}
This section describes the subset of the public CLERC \citep{hou2024clerc} and BriefMe \citep{woo2025briefme} datasets used in this study. Statutes, regulations, and other non-case authorities are outside the scope of this work, so we filter out citations to those authorities. We first report the composition of the retained citations, focusing on whether they include pinpoint page references and on the functional role they play in the surrounding text. We then evaluate the prompt-based classifier used to identify substantive citations.

\paragraph{Dataset Composition.}
Tables \ref{tab:pinpoint-stats} and \ref{tab:citation-type-stats} report the composition of the retained case law citations along two dimensions: whether citations include pinpoint page references and the functional role each citation plays under our taxonomy.

\begin{table}[h]
\centering
\small
\setlength{\tabcolsep}{8pt}
\begin{tabular}{@{}lrr@{}}
\softtoprule
 & \textbf{CLERC} & \textbf{BriefMe} \\
\midrule
Total citations & $2,965$ & $641$ \\
\midrule
Pinpoint & $1,074$ ($36\%$) & $145$ ($23\%$) \\
Non-pinpoint & $1,891$ ($64\%$) & $496$ ($77\%$) \\
\bottomrule
\end{tabular}
\captionsetup{width=0.7\textwidth}
\caption{Pinpoint coverage among retained case law citations. Pinpoint citations give a specific page or page range within the cited case, such as ``528 U.S. 440, 442'' or ``482 U.S. 451, 462--63.'' They are required for the Hard corruption setting, where we keep the cited case fixed and alter only the cited page.}
\label{tab:pinpoint-stats}
\end{table}

\begin{table}[h]
\centering
\small
\setlength{\tabcolsep}{8pt}
\begin{tabular}{@{}lrr@{}}
\softtoprule
\textbf{Citation Type} & \textbf{CLERC} & \textbf{BriefMe} \\
\midrule
Substantive & $2,064$ ($70\%$) & $442$ ($69\%$) \\
Procedural & $319$ ($11\%$) & $25$ ($4\%$) \\
Secondary & $582$ ($20\%$) & $170$ ($27\%$) \\
\bottomrule
\end{tabular}
\captionsetup{width=0.7\textwidth}
\caption{Citation-role distribution under our three-category taxonomy. Substantive citations directly support legal propositions and form the evaluation set. Procedural citations record litigation history, while secondary citations provide background, comparison, or contrary authority.}
\label{tab:citation-type-stats}
\end{table}

We make the following observations. First, non-pinpoint citations are more common in BriefMe than in CLERC. One plausible reason is that legal briefs often distribute support across statutes and cases, while our evaluation keeps only case law citations. Court opinions, by contrast, more often cite specific pages when justifying holdings. These pinpoint citations enable our Hard corruption strategy, in which changing only the page number creates subtle but meaningful errors.

Second, substantive citations dominate both datasets at similar rates ($70\%$ in CLERC and $69\%$ in BriefMe), confirming that both document types primarily use citations as direct support for legal propositions rather than for procedural history or background. Procedural citations are more common in CLERC ($11\%$) than in BriefMe ($4\%$), which is expected because court opinions often recount how a dispute moved through the courts. BriefMe contains a larger share of secondary citations ($27\%$) than CLERC ($20\%$), likely because briefs more frequently distinguish or compare prior authority when arguing a position.

\paragraph{Citation Classification Performance.}
Because our evaluation focuses on citations that directly support legal propositions, we first classify each retained citation under the three-category taxonomy introduced in \cref{sec:taxonomy}. We use GPT-4.1-mini with a simple prompt that has placeholders for the source passage, cited-document context, and taxonomy definitions. To evaluate this prompt-based classifier, we compare its predictions with labels from a held-out CLERC set of approximately $80$ examples annotated by the same three legal experts who developed the taxonomy. \Cref{tab:classification-perf} reports performance on the combined CLERC validation set and on the Medium and Hard subsets.

\begin{table}[h]
\centering
\small
\setlength{\tabcolsep}{6pt}
\begin{tabular}{@{}lcccc@{}}
\softtoprule
\textbf{Evaluation Setting} & \textbf{F1} & \textbf{Prec.} & \textbf{Rec.} & \textbf{Acc.} \\
\midrule
Combined (all) & $0.83$ & $0.83$ & $0.83$ & $0.83$ \\
Medium & $0.79$ & $0.79$ & $0.79$ & $0.79$ \\
Hard & $0.81$ & $0.81$ & $0.81$ & $0.81$ \\
\bottomrule
\end{tabular}
\caption{Validation performance of the prompt-based citation-type classifier on CLERC. GPT-4.1-mini assigns each retained citation to the taxonomy in \cref{sec:taxonomy}. Scores are computed against legal expert annotations on the combined validation set and separately on the Medium and Hard subsets. BriefMe is omitted from this breakdown because it does not contain a Medium split.}
\label{tab:classification-perf}
\end{table}

\newpage
\section{Human Annotation Details}
\label{appendix:human_annotation}

Here, we describe the details of the human annotation study presented in \cref{sec:human-validation}. Three legal analysts with prior litigation experience and expertise in U.S. federal case law, who had previously helped design the citation taxonomy, annotated a small subset of examples from the two datasets. Analysts were presented with the source passage (with the citation marked) and the full cited document, and were asked to label whether the citation was ``on point'' or ``not on point.'' Each annotator evaluated $23$ citations per dataset; $10$ of these were rated by all three annotators. These $10$ citations were used to measure inter-annotator agreement. The annotation samples included both on-point (valid) and not-on-point (corrupted) examples, so annotators could not trivially mark all examples with the same label.

\paragraph{Annotation Guidelines.}
Annotators were instructed to consider a citation ``on point'' if the cited authority supported the legal proposition in the source passage. In making this decision, the instructions directed annotators to consider proposition-level support rather than mere topical relevance.

\paragraph{Extended Agreement Metrics.}
\Cref{tab:iaa-extended} shows additional agreement metrics, including Fleiss' $\kappa$. For BriefMe, Fleiss' $\kappa$ is low ($-0.07$) due to class imbalance (prevalence paradox), while Gwet's AC1 remains high ($0.85$), indicating genuine agreement. Traditional agreement metrics such as Fleiss' $\kappa$ can be misleading for skewed distributions typical of annotation tasks; Gwet's AC1 provides a more robust measure in such cases \citep{pradhan2025llmjudge}.

\begin{table}[h]
\centering
\small
\begin{tabular}{lccc}
\softtoprule
\textbf{Dataset} & \textbf{3-way Agree} & \textbf{Fleiss' $\kappa$} & \textbf{Gwet's AC1} \\
\midrule
CLERC & $90\%$ & $0.86$ & $0.88$ \\
BriefMe & $80\%$ & $-0.07$ & $0.85$ \\
\bottomrule
\end{tabular}
\caption{Extended inter-annotator agreement metrics.}
\label{tab:iaa-extended}
\end{table}

\paragraph{Agreement by Difficulty Level.}
Our annotation sample was not stratified by difficulty, resulting in uneven coverage across corruption types. For CLERC, $7$ of the $10$ overlapping cases were Hard corruptions, showing $86\%$ three-way agreement (AC1 = $0.81$). For BriefMe, $7$ of the $10$ were Medium corruptions, also showing $86\%$ three-way agreement. Easy and Positive categories had insufficient overlap for reliable estimation. The strong agreement on Medium and Hard examples, where we expected the most annotator uncertainty, suggests that our corruptions produce clear ``not on point'' examples even for the pinpoint-page corruptions.

\paragraph{Agreement with Anticipated Labels.}
The preceding analyses measure agreement among annotators. We also ask whether annotators agree with the labels implied by our construction: ``on point'' for original citations and ``not on point'' for corrupted citations. \Cref{tab:label-agreement} reports, for each annotator, the share of examples whose human label matches this anticipated label. High agreement indicates that experts generally judged the original citations as valid and the corrupted citations as not on point, supporting the validity of our corruption procedure.

\begin{table}[h]
\centering
\small
\begin{tabular}{lcccc}
\softtoprule
\textbf{Dataset} & \textbf{Ann. 1} & \textbf{Ann. 2} & \textbf{Ann. 3} & \textbf{Avg.} \\
\midrule
CLERC & $95.7\%$ & $91.7\%$ & $77.3\%$ & $88.2\%$ \\
BriefMe & $91.7\%$ & $91.3\%$ & $69.6\%$ & $84.2\%$ \\
\bottomrule
\end{tabular}
\caption{Annotator agreement with anticipated (heuristic) labels.}
\label{tab:label-agreement}
\end{table}

\newpage
\section{Error Analysis}
\label{appendix:error_analysis}

We examine GPT-5's actual rationales on Hard cases where it incorrectly labeled corrupted citations as ``on point.'' These are real model outputs, not constructed examples. \Cref{tab:error-examples} shows representative failures.

\begin{table}[h]
\centering
\small
\begin{tabular}{p{0.30\textwidth}p{0.50\textwidth}p{0.12\textwidth}}
\softtoprule
\textbf{Citation \& Context} & \textbf{GPT-5 Rationale (Verbatim, Incorrect)} & \textbf{Error Type} \\
\midrule
\textit{SEC v. Chenery Corp.}, 63 S.Ct. 454, at \textcolor{red}{457} (correct: 462). \newline
Context: courts should not require agencies to justify discretion ``in any particular manner.'' &
``\textit{Chenery} (1943) states a reviewing court is not suggesting that an agency must justify its exercise of administrative discretion `in any particular manner or with artistic refinement.' \textbf{The target content includes this exact language}, directly supporting the source paragraph's assertion.'' &
Fabricated verification \\
\midrule
\textit{Gross v. FBL Fin. Servs.}, 129 S.Ct. 2343, at \textcolor{red}{2355} (correct: 2352). \newline
Context: ADEA plaintiff must prove age was ``but-for'' cause. &
``The target content \textbf{expressly states} this holding (`a plaintiff\ldots\ must prove\ldots\ that age was the but-for cause of the challenged adverse employment action'), \textbf{directly matching} the source paragraph's proposition.'' &
Fabricated verification \\
\midrule
\textit{In re Donald J. Trump Casino Sec. Litig.}, 7 F.3d 357, at \textcolor{red}{371} (correct: 364). \newline
Context: bespeaks caution doctrine means ``statements must be analyzed in context.'' &
``The case \textbf{explicitly says}, `we must consider an alleged misrepresentation within the context in which it was communicated'\ldots\ The citation is on point.'' &
Fabricated verification \\
\bottomrule
\end{tabular}
\caption{Actual GPT-5 rationales on Hard cases where it incorrectly accepted corrupted citations. In each case, the model claims the cited page contains specific text supporting the proposition. Bold text highlights the model's verification claims. The model does not merely fail to check---it asserts, incorrectly, that it \textit{has} checked and found supporting content.}
\label{tab:error-examples}
\end{table}

\paragraph{Patterns in Model Failures.}
Two patterns emerge from the error analysis:

\begin{enumerate}
\item \textbf{Fabricated verification.} In roughly two-thirds of false negatives, models do not merely accept the citation passively; they claim the cited page ``expressly states,'' ``explicitly says,'' or ``directly supports'' the proposition. The model constructs a plausible-sounding legal rationale asserting it verified content that does not appear on the cited page. A practitioner reading such a rationale would have no reason to doubt it.

\item \textbf{Topical matching in place of propositional verification.} Because Hard corruptions swap pages within the same case, the target content is always topically related: same dispute, often the same statute or doctrine. The model treats this overlap as sufficient rather than checking whether the page states the proposition being cited. This explains the Easy-to-Hard gap: Easy corruptions substitute a different case entirely, making topical mismatch obvious.
\end{enumerate}

\paragraph{Verbatim Quotes as a Stress Test.}
Many legal passages quote the cited authority verbatim, providing an objective test: does the quoted text appear on the cited page? Among GPT-5's false negatives on Hard cases where the source paragraph contains a verbatim quote, the quoted language is absent from the cited page $92\%$ of the time. If the model were checking whether the page contains the language it is supposed to support, these cases would be straightforward.

\Cref{tab:quote-examples} shows representative examples. In each case, the source paragraph quotes specific language and attributes it to a page that does not contain that language.

\begin{table}[h]
\centering
\small
\begin{tabular}{p{0.30\textwidth}p{0.32\textwidth}p{0.30\textwidth}}
\softtoprule
\textbf{Passage with Quote} & \textbf{Wrong Page Content} & \textbf{GPT-5 Rationale} \\
\midrule
\textit{In re Trump Casino}, 7 F.3d 357, at \textcolor{red}{371} (correct: 364). \newline
Passage quotes: ``merely reflects the unremarkable proposition that statements must be analyzed in context.'' &
Page $371$ discusses casino competition in Atlantic City: ``Competition in the Atlantic City casino/hotel market is intense. At present, there are twelve casino/hotels\ldots'' &
``The case \textbf{explicitly says}, `we must consider an alleged misrepresentation within the context in which it was communicated.'\,'' \newline
\textit{[This text does not appear on page $371$.]} \\
\midrule
\textit{Technical Res. Servs. v. Dornier}, 134 F.3d 1458, at \textcolor{red}{1467} (correct: 1464). \newline
Passage quotes: ``it is the duty of the courts to attempt to harmonize the answers.'' &
Page $1467$ discusses antitrust business justifications: ``the jury chose to credit some or all of DMSI's business justifications\ldots'' &
``The target content \textbf{expressly holds} that\ldots\ the court must adopt [a consistent] view.'' \newline
\textit{[Page $1467$ contains no such holding.]} \\
\midrule
\textit{United States v. Seckinger}, 90 S.Ct. 880, at \textcolor{red}{890--891} (correct: 884--885). \newline
Passage quotes: ``construe the contract language most strongly against the drafter.'' &
Page $890$--$891$ is a dissent discussing government contract clause interpretation, not the contra proferentem rule. &
``The cited passage \textbf{emphasizes} that\ldots\ ambiguous standard-form language will not be read to shift liability---reflecting contra proferentem.'' \newline
\textit{[The dissent does not state this rule.]} \\
\bottomrule
\end{tabular}
\caption{Cases where the source passage quotes the cited authority verbatim, but the corrupted page does not contain the quoted text. GPT-5 still accepts the citation and falsely claims to have found supporting content. Bold text highlights the model's false verification claims.}
\label{tab:quote-examples}
\end{table}

\newpage
\section{Evaluation Data Format}
\label{appendix:dataset_format}

We store evaluation examples as JSONL (JSON Lines) files; each line contains a single example. The ground truth is determined by the filename:

\begin{itemize}
    \item \texttt{*-on\_point.jsonl} --- Valid citations that correctly support the claimed proposition.
    \item \texttt{*-not\_on\_point-easy.jsonl} --- Easy: citation replaced with one from an unrelated document.
    \item \texttt{*-not\_on\_point-medium.jsonl} --- Medium: citation replaced with one from the same source document but different section.
    \item \texttt{*-not\_on\_point-hard.jsonl} --- Hard: only the pinpoint page number changed within the same cited case.
\end{itemize}

\noindent In our evaluation, ``not on point'' is treated as the positive class for detection, since the goal is to identify erroneous citations.

\subsection{Processing Beyond Source Datasets}
\label{appendix:contributions}

Our evaluation builds upon two existing legal NLP resources. We describe what each provides and what processing we apply.

\paragraph{CLERC.} The original CLERC dataset \citep{hou2024clerc} provides court opinion text with citations as (citation string, full cited document) pairs for case retrieval and retrieval-augmented generation. We extract individual citations with character-level positions, distinguish pinpoint citations (e.g., ``500 U.S. at 105'') from non-pinpoint citations, and extract page content for pinpoint citations into the \texttt{tgt\_paragraph} field. We classify citations by type (substantive, procedural, secondary) and apply the three-level corruption pipeline.

\paragraph{BriefMe.} The original BriefMe dataset \citep{woo2025briefme} provides Supreme Court brief texts with section headers for argument summarization and completion tasks. We extract citations from brief text, retrieve the full text of cited judicial opinions, identify pinpoint citations, extract their page content, and restructure everything into the verification task format with the \texttt{ref\_citations} array. We apply Easy and Hard corruptions; Medium is unavailable due to document structure differences.

\subsection{Schema Definition}
\label{appendix:schema}

Each evaluation example follows the structure shown below. A paragraph may contain multiple case law citations; we extract those classified as \textbf{substantive} (directly supporting legal propositions) and include them in the \texttt{ref\_citations} list.

\begin{tcolorbox}[title=JSONL Record Schema,
    colback=gray!5,
    colframe=gray!70!black,
    fonttitle=\bfseries,
    boxrule=1.5pt,
    arc=4pt,
    width=\textwidth,
    left=3mm,
    right=3mm]
\begin{lstlisting}[style=promptstyle]
{
  "src_paragraph": <string>,
  "src_document": <string>,              // CLERC only
  "ref_citations": [
    {
      "citation": <string>,
      "original_citation": <string>,     // corrupted files only
      "startPosition": <int>,
      "endPosition": <int>,
      "is_pinned_citation": <boolean>,
      "citation_on_point_overall": <string>,
      "tgt_document": <string>,
      "tgt_paragraph": <string>          // pinpoint citations only
    }
  ]
}
\end{lstlisting}
\end{tcolorbox}

\begin{table}[h]
\centering
\small
\begin{tabular}{>{\ttfamily}l p{0.55\textwidth}}
\softtoprule
\textbf{Field} & \textbf{Description} \\
\midrule
src\_paragraph & A paragraph from a legal document containing one or more case law citations. Substantive citations are extracted and placed in \texttt{ref\_citations}. \\
\addlinespace
src\_document & (CLERC only) Preceding paragraphs from the source document, providing additional context. Not present in BriefMe. \\
\addlinespace
ref\_citations & List of substantive citation objects extracted from the paragraph. \\
\midrule
\multicolumn{2}{l}{\textit{Fields within each \texttt{ref\_citations} entry:}} \\
\addlinespace
citation & The citation string as it appears in the (possibly corrupted) paragraph. \\
\addlinespace
original\_citation & (Corrupted files only) The citation before corruption was applied. Compare with \texttt{citation} to see what changed. \\
\addlinespace
startPosition & Character offset in \texttt{src\_paragraph} where the citation text begins. \\
\addlinespace
endPosition & Character offset where the citation ends. \texttt{src\_paragraph[startPosition:endPosition]} equals the citation text. \\
\addlinespace
is\_pinned\_citation & Boolean: true if the citation includes a pinpoint page reference (e.g., ``at 1467''). \\
\addlinespace
citation\_on\_point\_overall & \texttt{"\_yes\_"} or \texttt{"\_no\_"}. For corrupted files, this reflects the corrupted citation's validity. \\
\addlinespace
tgt\_document & The full text of the cited document. \\
\addlinespace
tgt\_paragraph & (Pinpoint citations only) The content from the specific cited page(s). Present when \texttt{is\_pinned\_citation} is true. \\
\bottomrule
\end{tabular}
\caption{Field definitions for the evaluation data schema. Ground truth is determined by filename.}
\label{tab:schema-fields}
\end{table}

\newpage
\subsection{Example: BriefMe Easy Corruption (Incorrect Prediction)}
\label{appendix:example_briefme}

This example shows an \textbf{Easy} corruption where the citation was replaced with one from an entirely unrelated case. The model incorrectly predicted ``on point.''

\begin{nolinenumbers}
\begin{tcolorbox}[title=BriefMe Example: Easy Corruption (Incorrect Prediction),
    colback=orange!5,
    colframe=orange!70!black,
    fonttitle=\bfseries,
    boxrule=1.5pt,
    arc=4pt,
    width=\textwidth,
    left=3mm,
    right=3mm]
\begin{lstlisting}[style=promptstyle]
{
  "src_paragraph": "...it would follow that neither state
    courts nor federal courts could review congressional
    redistricting plans... Yet 556 U.S. 1, held that
    'nothing in the language of [the Elections Clause]
    gives support to a construction that would immunize
    state congressional apportionment laws... from the
    power of courts to protect the constitutional rights
    of individuals.'",

  "ref_citations": [{
    "citation": "556 U.S. 1",
    "original_citation": "376 U.S. 1",
    "is_pinned_citation": false,
    "citation_on_point_overall": "_no_",
    "tgt_document": "KENNEDY, J., announced the judgment
      of the Court... This case requires us to interpret
      Section 2 of the Voting Rights Act of 1965..."
  }]
}
\end{lstlisting}
\end{tcolorbox}
\end{nolinenumbers}

\paragraph{Corruption Applied.} The original citation \textbf{376 U.S. 1} (\textit{Wesberry v. Sanders}, the landmark ``one person, one vote'' case about Elections Clause limits) was replaced with \textbf{556 U.S. 1} (\textit{Bartlett v. Strickland}, a Voting Rights Act Section 2 case about majority-minority districts).

\paragraph{Why Not On Point.} The paragraph quotes language about the Elections Clause and courts' power to review redistricting. This quote comes from \textit{Wesberry}, not \textit{Bartlett}. \textit{Bartlett} addresses an entirely different legal question (VRA Section 2 requirements) and does not contain the quoted language.

\paragraph{Model Error.} The model predicted ``YES'' (on point). Both cases involve voting rights and redistricting, so the topic matches. But topical similarity is not enough. The specific legal holding and quoted language do not appear in \textit{Bartlett}.

\newpage
\subsection{Example: CLERC Hard Corruption (Incorrect Prediction)}
\label{appendix:example_clerc}

This example shows a \textbf{Hard} corruption where only the pinpoint page number was changed. The model incorrectly predicted ``on point.''

\begin{nolinenumbers}
\begin{tcolorbox}[title=CLERC Example: Hard Corruption (Model Incorrect),
    colback=red!5,
    colframe=red!70!black,
    fonttitle=\bfseries,
    boxrule=1.5pt,
    arc=4pt,
    width=\textwidth,
    left=3mm,
    right=3mm]
\begin{lstlisting}[style=promptstyle]
{
  "src_document": "[preceding paragraphs from court opinion]",

  "src_paragraph": "The verdict on the King slot therefore
    establishes that the interviews... did not evidence
    discrimination as to the Dade slot either. See Technical
    Res. Servs., Inc. v. Dornier Med. Sys., Inc., 134 F.3d
    1458, 1467 (11th Cir.1998) ('[I]t is the duty of the
    courts to attempt to harmonize the answers, if it is
    possible under a fair reading of them...' (quoting
    Gallick v. Baltimore & Ohio R.R. Co., 372 U.S. 108)).",

  "ref_citations": [{
    "citation": "134 F.3d 1458, 1467",
    "original_citation": "134 F.3d 1458, 1464",
    "startPosition": 142,
    "endPosition": 161,
    "is_pinned_citation": true,
    "citation_on_point_overall": "_no_",
    "tgt_document": "[full text of Technical Res. Servs.
      v. Dornier Med. Sys.]",
    "tgt_paragraph": "A fair and reasonable reading of the
      jury's verdict is that the jury chose to credit some
      or all of DMSI's business justifications... DMSI also
      asserted: (1) concerns about product liability
      exposure, (2) a desire to guarantee quality..."
  }]
}
\end{lstlisting}
\end{tcolorbox}
\end{nolinenumbers}

\paragraph{Corruption Applied.} The pinpoint page was changed from $\mathbf{1464}$ to $\mathbf{1467}$ within the same case (\textit{Technical Res. Servs. v. Dornier Med. Sys.}).

\paragraph{Why Not On Point.} The paragraph quotes language about courts' duty to harmonize jury answers, citing page $1467$. But page $1467$ of \textit{Technical Res. Servs.} discusses antitrust issues and DMSI's business justifications. It says nothing about harmonizing jury verdicts. The actual quote appears on page $1464$.

\paragraph{Model Error.} The model predicted ``YES'' (on point). It correctly identified that the case involves a jury verdict, but failed to verify that page $1467$ contains the specific legal principle being cited. The model performed \textit{case-level} verification (correct case, mentions juries) rather than \textit{page-level} verification (does this page contain the quoted principle?).

\newpage
\section{Full Prompt Sensitivity Results}
\label{appendix:ablation_full}

This section provides complete results for the prompt intervention analysis in \cref{sec:prompt_intervention}, including Easy and Medium recall that are omitted from the main tables.

\subsection{CLERC (Court Opinions)}
\label{appendix:ablation_clerc}

\begin{table}[h!]
\centering
\small
\begin{tabular}{@{}lcccccccc@{}}
\softtoprule
 & \multicolumn{2}{c}{\textbf{Easy (\%)}} & \multicolumn{2}{c}{\textbf{Medium (\%)}} & \multicolumn{2}{c}{\textbf{Hard (\%)}} & \multicolumn{2}{c}{\textbf{FPR (\%)}} \\
\textbf{Model} & \textbf{Base} & \textbf{+Gnd.} & \textbf{Base} & \textbf{+Gnd.} & \textbf{Base} & \textbf{+Gnd.} & \textbf{Base} & \textbf{+Gnd.} \\
\midrule
GPT-4.1 & $97.6$ & $99.8$\pmark{+2.2} & $73.6$ & $90.9$\pmark{+17.3} & $38.4$ & $66.2$\pmark{+27.8} & $11.2$ & $22.7$\nmark{+11.5} \\
GPT-4.1-mini & $94.8$ & $97.5$\pmark{+2.7} & $76.3$ & $84.4$\pmark{+8.1} & $42.0$ & $58.7$\pmark{+16.7} & $20.0$ & $28.0$\nmark{+8.0} \\
GPT-4o & $99.8$ & $100.0$\pmark{+0.2} & $91.6$ & $97.9$\pmark{+6.3} & $60.6$ & $77.5$\pmark{+16.9} & $34.9$ & $49.3$\nmark{+14.4} \\
GPT-4o-mini & $98.9$ & $99.8$\pmark{+0.9} & $84.8$ & $96.4$\pmark{+11.6} & $59.5$ & $82.6$\pmark{+23.1} & $43.7$ & $64.2$\nmark{+20.5} \\
GPT-5 & $99.6$ & $99.8$\pmark{+0.2} & $87.5$ & $93.7$\pmark{+6.2} & $55.5$ & $68.8$\pmark{+13.3} & $11.6$ & $16.9$\nmark{+5.3} \\
GPT-5-mini & $96.2$ & $98.0$\pmark{+1.8} & $74.6$ & $85.9$\pmark{+11.3} & $43.6$ & $63.4$\pmark{+19.8} & $9.8$ & $16.4$\nmark{+6.6} \\
GPT-5.4 & $99.6$ & $99.9$\pmark{+0.3} & $90.2$ & $95.2$\pmark{+5.0} & $48.5$ & $66.3$\pmark{+17.8} & $15.6$ & $22.6$\nmark{+7.0} \\
GPT-5.4 (reasoning) & $99.8$ & $99.9$\pmark{+0.1} & $90.3$ & $94.9$\pmark{+4.6} & $59.8$ & $73.2$\pmark{+13.4} & $13.1$ & $19.5$\nmark{+6.4} \\
\midrule
Claude Sonnet 4 & $99.2$ & $99.7$\pmark{+0.5} & $88.6$ & $93.2$\pmark{+4.6} & $47.9$ & $57.6$\pmark{+9.7} & $12.8$ & $18.8$\nmark{+6.0} \\
Claude Sonnet 4.6 & $99.8$ & $99.8$\pmark{+0.0} & $93.0$ & $96.5$\pmark{+3.5} & $44.3$ & $62.0$\pmark{+17.7} & $7.7$ & $14.0$\nmark{+6.3} \\
Claude Opus 4.6 & $99.2$ & $99.6$\pmark{+0.4} & $81.2$ & $87.2$\pmark{+6.0} & $36.5$ & $51.5$\pmark{+15.0} & $4.1$ & $5.1$\nmark{+1.0} \\
\midrule
Gemini 2.5 Flash & $98.9$ & $99.6$\pmark{+0.7} & $83.2$ & $93.1$\pmark{+9.9} & $59.2$ & $77.9$\pmark{+18.7} & $11.9$ & $36.4$\nmark{+24.5} \\
Gemini 2.5 Pro & $99.5$ & $99.6$\pmark{+0.1} & $84.4$ & $88.7$\pmark{+4.3} & $58.3$ & $72.9$\pmark{+14.6} & $11.9$ & $19.5$\nmark{+7.6} \\
Gemini 3.1 Pro & $99.7$ & $99.8$\pmark{+0.1} & $88.5$ & $94.2$\pmark{+5.7} & $56.3$ & $67.7$\pmark{+11.4} & $11.0$ & $13.7$\nmark{+2.7} \\
\bottomrule
\end{tabular}
\caption{Full CLERC results: baseline prompt (Base) vs.\ page-grounded prompt (+Gnd.) across all difficulty levels. Easy and Medium recall show modest improvements; Hard recall shows the largest gains.}
\label{tab:ablation-clerc-full}
\end{table}

\subsection{BriefMe (Legal Briefs)}
\label{appendix:ablation_briefme}

\begin{table}[h!]
\centering
\small
\begin{tabular}{@{}lcccccc@{}}
\softtoprule
 & \multicolumn{2}{c}{\textbf{Easy (\%)}} & \multicolumn{2}{c}{\textbf{Hard (\%)}} & \multicolumn{2}{c}{\textbf{FPR (\%)}} \\
\textbf{Model} & \textbf{Base} & \textbf{+Gnd.} & \textbf{Base} & \textbf{+Gnd.} & \textbf{Base} & \textbf{+Gnd.} \\
\midrule
GPT-4.1 & $96.2$ & $99.6$\pmark{+3.4} & $51.5$ & $87.0$\pmark{+35.5} & $16.1$ & $30.8$\nmark{+14.7} \\
GPT-4.1-mini & $93.0$ & $97.4$\pmark{+4.4} & $59.2$ & $76.0$\pmark{+16.8} & $19.5$ & $30.6$\nmark{+11.1} \\
GPT-4o & $99.4$ & $100.0$\pmark{+0.6} & $79.6$ & $89.0$\pmark{+9.4} & $50.8$ & $59.4$\nmark{+8.6} \\
GPT-4o-mini & $99.1$ & $99.6$\pmark{+0.5} & $71.8$ & $93.0$\pmark{+21.2} & $52.0$ & $71.7$\nmark{+19.7} \\
GPT-5 & $100.0$ & $99.8$\pmark{-0.2} & $77.0$ & $91.0$\pmark{+14.0} & $14.4$ & $22.5$\nmark{+8.1} \\
GPT-5-mini & $96.2$ & $98.9$\pmark{+2.7} & $59.0$ & $89.0$\pmark{+30.0} & $13.9$ & $26.7$\nmark{+12.8} \\
GPT-5.4 & $97.7$ & $99.8$\pmark{+2.1} & $64.0$ & $91.0$\pmark{+27.0} & $14.6$ & $24.2$\nmark{+9.6} \\
GPT-5.4 (reasoning) & $99.4$ & $99.6$\pmark{+0.2} & $82.0$ & $89.0$\pmark{+7.0} & $14.4$ & $21.1$\nmark{+6.7} \\
\midrule
Claude Sonnet 4 & $99.4$ & $99.8$\pmark{+0.4} & $68.6$ & $85.0$\pmark{+16.4} & $28.5$ & $38.5$\nmark{+10.0} \\
Claude Sonnet 4.6 & $99.4$ & $99.8$\pmark{+0.4} & $64.6$ & $90.7$\pmark{+26.1} & $18.6$ & $24.4$\nmark{+5.8} \\
Claude Opus 4.6 & $97.2$ & $97.7$\pmark{+0.5} & $54.0$ & $77.3$\pmark{+23.3} & $5.4$ & $7.1$\nmark{+1.7} \\
\midrule
Gemini 2.5 Flash & $97.7$ & $99.4$\pmark{+1.7} & $74.8$ & $91.0$\pmark{+16.2} & $12.0$ & $36.7$\nmark{+24.7} \\
Gemini 2.5 Pro & $99.1$ & $99.3$\pmark{+0.2} & $82.7$ & $90.2$\pmark{+7.5} & $16.3$ & $28.6$\nmark{+12.3} \\
Gemini 3.1 Pro & $99.6$ & $100.0$\pmark{+0.4} & $82.7$ & $89.2$\pmark{+6.5} & $11.9$ & $19.0$\nmark{+7.1} \\
\bottomrule
\end{tabular}
\caption{Full BriefMe results: baseline prompt (Base) vs.\ page-grounded prompt (+Gnd.). BriefMe lacks the Medium difficulty level.}
\label{tab:ablation-briefme-full}
\end{table}

\newpage
\section{Diagnostic Results on Distinguished and Criticized Authorities}
\label{appendix:relational_results}

Legal opinions can refer to other opinions in several ways. These references are often described in terms of how one case treats another. A case may \textit{distinguish} an earlier authority, meaning it explains why that authority does not apply because the facts or legal posture differ. A case may also \textit{criticize} another authority, meaning it questions or rejects some part of that authority's reasoning without necessarily overruling it.

We use these treatment relations to construct a small diagnostic test set. We ask: when two authorities are legally related, that is, when one distinguishes or criticizes the other, do LLMs still separate on-point citations from not-on-point citations as expected, or does the relation itself make this check more difficult?

We constructed four small datasets around this question. Two are valid sets: they contain on-point citations where the cited authority happens to be involved in a distinguishing or criticizing relation. In these examples, the source paragraph still cites the correct case, so the citation remains propositionally valid.

The other two datasets consist of not-on-point examples. We start with a paragraph that correctly cites one case. We then look inside that cited case for another citation, where the cited case distinguishes or criticizes a different authority. Finally, we replace the paragraph's original citation with this second citation. The replacement may look plausible because the substituted authority is only one citation hop away: it appears inside the originally cited case in a discussion where that case distinguishes or criticizes it. However, this relationship does not make the substituted citation on point. The paragraph was making a claim about the original case, not the substituted authority.

Across models and prompt versions, the evaluated sample sizes range from $24$--$27$ for the distinguished on-point set, $33$--$35$ for the distinguished not-on-point set, $17$--$18$ for the criticized on-point set, and $34$--$37$ for the criticized not-on-point set.

\Cref{tab:relational-replacements} reports false positive rate on the valid sets and recall on the not-on-point sets under both prompts, using the same ``All Citations'' evaluation protocol as in the main paper. The results suggest that treatment-relation replacements are challenging in two ways. First, several models over-reject valid citations when the cited authority appears in a passage that distinguishes or criticizes another authority. Second, page-grounded prompting generally improves recall on the not-on-point variants, but often at the cost of higher false positive rates on the valid variants, mirroring the trade-off seen in the main prompt intervention.

\begin{table}[h!]
\centering
\footnotesize
\begin{tabular}{@{}lcccccccc@{}}
\softtoprule
 & \multicolumn{2}{c}{\textbf{Dist. FPR (\%)}} & \multicolumn{2}{c}{\textbf{Dist. Recall (\%)}} & \multicolumn{2}{c}{\textbf{Crit. FPR (\%)}} & \multicolumn{2}{c}{\textbf{Crit. Recall (\%)}} \\
\textbf{Model} & \textbf{Base} & \textbf{+Gnd.} & \textbf{Base} & \textbf{+Gnd.} & \textbf{Base} & \textbf{+Gnd.} & \textbf{Base} & \textbf{+Gnd.} \\
\midrule
GPT-4.1 & $29.6$ & $32.0$ & $76.5$ & $82.9$ & $17.6$ & $33.3$ & $81.1$ & $86.5$ \\
GPT-4.1-mini & $29.6$ & $32.0$ & $70.6$ & $80.0$ & $17.6$ & $44.4$ & $81.1$ & $83.8$ \\
GPT-4o & $74.1$ & $84.0$ & $100.0$ & $100.0$ & $76.5$ & $88.9$ & $94.6$ & $100.0$ \\
GPT-4o-mini & $66.7$ & $96.0$ & $94.1$ & $100.0$ & $58.8$ & $88.9$ & $89.2$ & $100.0$ \\
GPT-5 & $14.8$ & $12.0$ & $91.2$ & $97.1$ & $0.0$ & $5.6$ & $73.0$ & $89.2$ \\
GPT-5-mini & $18.5$ & $24.0$ & $70.6$ & $94.3$ & $5.9$ & $5.6$ & $73.0$ & $91.9$ \\
GPT-5.4 & $14.8$ & $0.0$ & $73.5$ & $82.9$ & $5.9$ & $11.1$ & $73.0$ & $89.2$ \\
GPT-5.4 (reasoning) & $11.1$ & $8.0$ & $97.1$ & $100.0$ & $11.8$ & $11.1$ & $81.1$ & $97.3$ \\
\midrule
Claude Sonnet 4 & $25.9$ & $24.0$ & $85.3$ & $94.3$ & $5.9$ & $16.7$ & $89.2$ & $94.6$ \\
Claude Sonnet 4.6 & $26.9$ & $16.0$ & $93.9$ & $100.0$ & $0.0$ & $5.9$ & $91.9$ & $100.0$ \\
Claude Opus 4.6 & $3.7$ & $0.0$ & $73.5$ & $87.9$ & $0.0$ & $0.0$ & $67.6$ & $79.4$ \\
\midrule
Gemini 2.5 Flash & $11.1$ & $16.0$ & $79.4$ & $82.9$ & $5.9$ & $44.4$ & $70.3$ & $86.5$ \\
Gemini 2.5 Pro & $7.4$ & $12.0$ & $88.2$ & $97.1$ & $0.0$ & $5.6$ & $70.3$ & $86.5$ \\
Gemini 3.1 Pro & $0.0$ & $0.0$ & $91.2$ & $94.3$ & $0.0$ & $0.0$ & $81.1$ & $89.2$ \\
\bottomrule
\end{tabular}
\caption{Exploratory results on distinguished and criticized citations. \textbf{Dist. FPR} and \textbf{Crit. FPR} are false positive rates on valid citations in the distinguished and criticized on-point sets. \textbf{Dist. Recall} and \textbf{Crit. Recall} are recall on the corresponding not-on-point sets. Results use the same baseline prompt (Base) and page-grounded prompt (+Gnd.) as the main paper.}
\label{tab:relational-replacements}
\end{table}

\begin{takeawaybox}
Substantive citations involving distinguished or criticized authorities are especially challenging for LLMs to assess. Such authorities are legally related to the source passage, but that relation does not by itself establish that the citation is on point.
\end{takeawaybox}

\newpage
\section{Primer on U.S. Legal Citation and Document Types}
\label{appendix:legal_background}

Here, we provide a brief overview of the U.S. legal documents and citation terms used throughout the paper. U.S. legal writing follows a specialized citation convention commonly known as Bluebook style. \textit{The Bluebook} is the standard guide to that system. The book provides lawyers with shared conventions for identifying sources and showing how those sources are used. The name `Bluebook' comes from the blue cover of the earlier editions of the book.

There are two important ideas to keep in mind when reading this paper. First, a citation has a ``surface form,'' meaning the words and numbers printed on the page. Second, a citation may be introduced by a signal, such as ``see,'' ``cf.,'' or ``but see.'' A signal tells the reader how the cited source relates to the claim. It may support the claim directly, support it by analogy, or point the other way. The citation below labels the basic pieces.

{\small
\[
\underbracket[0.6pt][2pt]{\textcolor{teal}{\mathit{see}}}_{\textcolor{teal}{\text{signal}}}\quad
\underbracket[0.6pt][2pt]{\textcolor{blue!70!black}{217}}_{\textcolor{blue!70!black}{\text{volume}}}\quad
\underbracket[0.6pt][2pt]{\textcolor{purple!80!black}{\mathrm{F.R.D.}}}_{\textcolor{purple!80!black}{\text{reporter}}}\quad
\underbracket[0.6pt][2pt]{\textcolor{orange!80!black}{309}}_{\textcolor{orange!80!black}{\text{first page}}},\quad
\underbracket[0.6pt][2pt]{\textcolor{red!70!black}{322}}_{\textcolor{red!70!black}{\text{pinpoint page}}}
\]
}

\subsection{Document Types}
\label{appendix:legal_background:documents}

We use the following document types and authorities throughout the paper.

\begin{itemize}[leftmargin=*,itemsep=2pt]
  \item \textbf{Court opinion}: a written judicial decision explaining how a court resolved a dispute. CLERC consists of court opinions.
  \item \textbf{Legal brief}: a filing written by lawyers to persuade a court. Briefs argue for a legal result and cite authority in support. BriefMe consists of legal briefs.
  \item \textbf{Case law}: the body of judicial decisions that lawyers and courts later cite as legal authority.
  \item \textbf{Statute}: enacted legislation, commonly cited by code title and section, such as ``42 U.S.C. \S 1983.'' We mention statutes for contrast, but exclude them from the evaluation.
  \item \textbf{Regulation}: an agency rule, often cited from the Code of Federal Regulations, such as ``17 C.F.R. \S 240.10b-5.'' We exclude these as well.
\end{itemize}

\subsection{Anatomy of a Case Citation}
\label{appendix:legal_background:anatomy}

A standard case citation identifies the case, the reporter in which it appears, and often the particular page being cited. In this paper, the basic pattern is:

\begin{quote}
\textit{Case Name}, Volume Reporter First Page, Pinpoint (Court Year)
\end{quote}

For example, consider the following citation.

\begin{quote}
\textit{Zubulake v. UBS Warburg LLC}, 217 F.R.D. 309, 322 (S.D.N.Y. 2003)
\end{quote}

It contains the following parts.

\begin{itemize}[leftmargin=*,itemsep=2pt]
  \item \textbf{Case name}: \textit{Zubulake v. UBS Warburg LLC}.
  \item \textbf{Volume}: 217.
  \item \textbf{Reporter}: \textit{F.R.D.}, the published series in which the opinion appears.
  \item \textbf{First page}: 309, where the case begins in that reporter.
  \item \textbf{Pinpoint page}: 322, the specific page offered in support of the proposition.
  \item \textbf{Court and year}: S.D.N.Y. 2003, identifying the court and decision year.
\end{itemize}

The pinpoint page is central to our task. A case may be relevant in general while the cited page fails to support the specific proposition in the surrounding text. That distinction motivates our Hard corruption setting and our phrase \textit{unsupported pinpoint citation}.

\subsection{Common Citation Patterns}
\label{appendix:legal_background:patterns}

Legal citations can appear in several recurring forms.

\begin{itemize}[leftmargin=*,itemsep=2pt]
  \item \textbf{Full case citation}: gives the case name, reporter, first page, and often a pinpoint page.
  \item \textbf{Short-form citation}: abbreviates a source after it has already been introduced, as in ``\textit{Mullane}, 339 U.S. at 314.''
  \item \textbf{Signal-led citation}: begins with a Bluebook signal such as \textit{see}, \textit{see also}, \textit{cf.}, or \textit{but see}. These signals tell the reader whether the cited source supports the proposition directly, supports it by analogy, or cuts against it.
  \item \textbf{Procedural-history citation}: records what happened earlier in a case, often with terms such as \textit{aff'd}, \textit{rev'd}, or \textit{vacated}.
  \item \textbf{Non-case citation}: cites a statute, regulation, treatise, or other authority rather than a judicial opinion. These citations are outside the scope of our evaluation.
\end{itemize}

\subsection{Other Terms Used in This Paper}
\label{appendix:legal_background:terms}

The following terms also recur throughout the paper.

\begin{itemize}[leftmargin=*,itemsep=2pt]
  \item \textbf{Authority}: the source offered in support of a legal claim.
  \item \textbf{Holding}: the legal determination or rule for which a case is cited.
  \item \textbf{On point}: a citation is on point when the cited authority supports the proposition for which it is offered.
  \item \textbf{Citechecking}: the process of checking citation form and verifying that the cited source supports the proposition in the text.
  \item \textbf{Substantive citation}: a citation offered as direct support for a legal proposition.
  \item \textbf{Procedural history citation}: a citation used to trace what happened earlier in the litigation, rather than to state a rule.
  \item \textbf{Secondary citation}: a citation used for background, comparison, or contrast, rather than as the main doctrinal support.
  \item \textbf{Binding authority}: authority that a court must follow.
  \item \textbf{Persuasive authority}: authority that a court may consider but need not follow.
  \item \textbf{Precedential reasoning}: legal argument that relies on prior cases and analogies to earlier decisions.
  \item \textbf{Unsupported pinpoint citation} or \textbf{pinpoint mismatch}: our shorthand for a citation where the case is real and the citation format is plausible, but the cited page does not support the proposition being asserted.
\end{itemize}

\end{document}